\documentclass[preprint2]{aastex}

\usepackage{graphicx}
\usepackage{natbib}
\usepackage{lscape}
\usepackage{rotating}
\usepackage{txfonts}

\newcommand{\kms}{km\ s$^{-1}$}

\newcommand{\xmm}{{\sc XMM}\emph{-Newton}}

\newcommand{\zp}{$\zeta$\,Puppis}

\begin{document}
%

  \title{A detailed X-ray investigation of \zp\ \\
The variability on short and long timescales\altaffilmark{5}}

\author{Ya\"el Naz\'e\altaffilmark{1,2}, Lidia M. Oskinova\altaffilmark{3}, and Eric Gosset\altaffilmark{2,4}}
\email{naze@astro.ulg.ac.be}

\altaffiltext{1}{Research Associate FRS-FNRS}
\altaffiltext{2}{GAPHE, D\'epartement AGO, Universit\'e de Li\`ege, All\'ee du 6 Ao\^ut 17, Bat. B5C, B4000-Li\`ege, Belgium}
\altaffiltext{3}{Institute for Physics and Astronomy, University of Potsdam, 14476 Potsdam, Germany}
\altaffiltext{4}{Senior Research Associate FRS-FNRS}
\altaffiltext{5}{Based on observations collected with XMM-Newton, an ESA Science Mission with instruments and contributions directly funded by ESA Member States and the USA (NASA).}

\begin{abstract}
Stellar winds are a crucial component of massive stars, but their exact properties still remain uncertain. To shed some light on this subject, we have analyzed an exceptional set of X-ray observations of \zp, one of the closest and brightest massive stars. The sensitive lightcurves that were derived reveal two major results. On the one hand, a slow modulation of the X-ray flux (with a relative amplitude of up to 15\% over 16h in the 0.3--4.0\,keV band) is detected. Its characteristic timescale cannot be determined with precision, but amounts from one to several days. It could be related to corotating interaction regions, known to exist in \zp\ from UV observations. Hour-long changes, linked to flares or to the pulsation activity, are not observed in the last decade covered by the XMM observations; the 17h tentative period, previously reported in a $ROSAT$ analysis, is not confirmed either and is thus transient, at best. On the other hand, short-term changes are surprisingly small ($<1$\% relative amplitude for the total energy band). In fact, they are compatible solely with the presence of Poisson noise in the data. This surprisingly low level of short-term variability, in view of the embedded wind-shock origin, requires a very high fragmentation of the stellar wind, for both absorbing and emitting features ($>10^5$ parcels, comparing with a 2D wind model). This is the first time that constraints have been placed on the number of clumps in an O-type star wind and from X-ray observations. 

\end{abstract}


%

\section{Introduction}
Stars more massive than 20 suns are the true `cosmic engines' of our Universe.  Through their winds and final explosions as supernovae, they shape the interstellar medium and largely contribute to its chemical enrichment. Besides, those stars are the most luminous ones, the only ones seen from afar. Studying distant galaxies or our close neighborhood thus requires a good knowledge of these objects. However, one of their most crucial physical properties, the stellar wind, remains poorly constrained.

In recent years a large debate arose in the massive star's community regarding the structure and strength of such winds. Based on the analysis of resonance lines in UV and FUV spectra, \citet{bou05} and \citet{ful06} revised mass-loss rates of a sample of O-type stars by orders of magnitude down compared to those obtained from H$\alpha$ and radio diagnostics. \citet{osk07} showed that including wind clumping in the analysis of UV resonance lines yielded mass-loss rates in agreement with those derived from H$\alpha$ (see also \citealt{sun11,sur12}).  These estimates crucially depend on the degree of wind clumping and exact properties of these clumps.

In this debate, X-ray observations may well play a crucial role. Indeed, the first high-resolution spectra of massive stars revealed line profiles which were more symmetric and less blueshifted than expected \citep[e.g.][]{kra03}, triggering (for a large part) the current mass-loss debate. The first global fitting of high-resolution X-ray spectra has just been attempted and it is showing the clumps to be rather spherical (and not ``pancake-like''), the non-porosity of the wind, and some of the plasma emitting components to be distributed over large regions \citep[ up to a hundred stellar radii in size, ][and Herv\'e et al., submitted]{her12}.

Another observable parameter, the variability, could also be used to learn about the wind structure. Indeed, the intrinsic instability of line-driven stellar winds generates clumps and shocks between clumps, which in turn produce high-energy emission over a large zone \citep{luc80,fel97b}. This ``embedded wind-shock model" currently is the standard model for X-ray production. Because the X-rays are directly born in the winds, their variability provides strong constraints on the wind structure. Hydrodynamical simulations showed the highly changing level of the produced X-rays, a variability which could only be lowered if many clumps were present \citep{fel97b}. A quantitative, theoretical assessment of the interplay between clump number and variation level, for different energy ranges, was performed only recently \citep{osk04}. 

On the observational side, searches for such high-energy variations were undertaken with several facilities, the most thorough work having been done with $ROSAT$ \citep{ber94b}, but no significant short-term variation was detected in O-stars. However, the sensitivity of these early observations was poor, with relative error bars in $ROSAT$ lightcurves amounting to several tens of percents, and they thus only provided weak constraints. 

In the last decade, \xmm\ has obtained an exceptional 1Ms dataset of a massive star, \zp. To date, there exists no more sensitive dataset, or a dataset with a better (time and spectral) coverage, for a massive star. We have thus undertaken the analysis of this dataset, with the aim of constraining the wind structure using X-ray variability. A first paper (\citealt{naz11}, hereafter Paper I) has presented these data, explained their reduction, and provided some first results (notably the analysis of EPIC spectra). This second paper focuses on the detailed variations recorded with EPIC and RGS instruments. It is organized as follows: the target is presented in Sect. 2, and the observations in Sect. 3; the observed changes found on long, intermediate, and short timescales are described in Sects. 4 to 6; the results are finally summarized in Sect. 7. An appendix provides details on the results for each exposure.

\section{\zp\ as a variable star}

Being one of the closest and brightest massive stars, \zp\ (O4Infp) provides the opportunity to perform an in-depth investigation of its properties, notably its variability. 

For example, \zp\ is well known for its spectral variations in the visible and UV domains. Some impressive variations have been reported, e.g. a 50\% decrease of H$\alpha$ strength over 2 years \citep{con76} and a doubling of the He{\sc ii}\,$\lambda$\,4686 and H$\alpha$ line profiles over successive nights \citep[i.e. from single line to double-peaked profiles,][]{weg78}. Periodic changes have also been detected, on short as well as longer timescales. A 5d modulation found in the UV and optical data (H$\alpha$ line, Si\,{\sc iv} doublet at 1400\AA, photometry in the Str\"omgren $b$ band) was interpreted as linked to the stellar rotation \citep{mof81, bal92, how95, ber96}. \citet{mof81} then further proposed that the inner wind regions are magnetically confined and forced to corotate with the star. Spectro-polarimetric observations did reveal an equatorial compression of the stellar wind of \zp\ \citep{har96} but failed to indicate the presence of a strong magnetic field \citep{sch08}. Shorter-term changes of the UV and optical line profiles with periods in the range 2--20h were also reported by several authors \citep[and references therein]{baa91,rei93,how95,ber96,rei96} and usually attributed to non-radial pulsations. Finally, \citet{eve98} detected stochastic substructures in the He\,{\sc ii}\,$\lambda$\,4686 emission line moving away from the line center with time, which they interpreted as the signature of small-scale clumps. 

In the X-ray domain too, the variability of \zp\ was investigated. Despite several early claims \citep{col89}, $Einstein$ observations did not reveal any significant variations: the false detections could be explained by instrumental effects \citep{sno81,ber96}. $ROSAT$ data, however, told another story \citep{ber96}. Variations of the mean count rate at high energies (0.9--2.0\,keV) were detected between the five $ROSAT$ exposures, but only at a 2\% significance level (i.e. lower than a 3$\sigma$ detection). The longest $ROSAT$ exposure (56.7ks spread over 11 days) was also analyzed to look for short-term variations. While $\chi^2$ and Kolmogorov-Smirnov methods did not yield definitive results, period folding techniques revealed the presence of a 1.44\,d$^{-1}$ frequency (=16.7h period) in the $ROSAT$ high-energy band (i.e. 0.9--2.0\,keV band). These high-energy variations have a small amplitude (about 6\%), hence the period is detected with only ``95\% confidence'', but an additional fact reinforced the confidence in their presence: the X-ray variations appeared in phase with changes in the H$\alpha$ profile. \citet{ber96} then proposed that the periodic modulation of the wind density in the lower wind layers (as traced by H$\alpha$) triggers wind instabilities which, in turn, produce wind shocks and therefore X-rays. \citet{ber96} however cautioned that their results should be checked with better quality X-ray data, and a later study of an $ASCA$ dataset of \zp\ did not confirm the previous result \citep{osk01b}. Using the best X-ray dataset available at the present time (18 \xmm\ exposures spread over a decade), we now try to investigate in detail the X-ray variability of \zp. 

\section{\xmm\ observations and their reduction}
In the past decade of \xmm\ observations, the star \zp\ was observed during 18 exposures, mostly for calibration purposes. These datasets are excellent for studying the variability of \zp\ since (1) the scheduled exposure times were often long (up to $\sim$70\,ks) and (2) the observing dates probe weekly, monthly, and yearly timescales. Unfortunately, many observations were affected by soft proton background flares, resulting in total exposure times reduced by about 30\%. A summary of the observations, as well as a detailed discussion of the reduction process (using SAS v10.0.0) and of the spectra generation are given in Paper I. 

Up to five instruments aboard \xmm\ simultaneously record the X-ray emission of sources (three with low spectral resolution: EPIC-pn, EPIC-MOS1, EPIC-MOS2; and two with higher spectral resolution: RGS1, RGS2). Because of their separate optical paths and detector characteristics, they can be considered as five independent instruments observing simultaneously. Indeed, the very different PSF observed with MOS1 and MOS2 is a clear example that no two instruments are identical aboard \xmm, even if their names are similar. Furthermore, each observation corresponds to a particular realization of the noise: therefore, it is possible (and even expected!) that each realization will slightly differ from another one. In the case of the \xmm\ observations of \zp, it means that a lightcurve recorded by one instrument will slightly differ from that recorded by another one, though they are compatible within the uncertainties. Having 5 independent datasets could be seen at first as a disadvantage, since any test or fitting will yield five not-exactly-identical answers, but that actually reinforces the conclusions. Indeed, it reminds us that looking at a single observation with a single instrument, as so often is done except when working with \xmm, does not allow any cross-checking and thus leaves room for spurious signal - in such cases, one must always be very careful not to over-interpret the data. Having multiple datasets enables us to check any detected signal, and in what follows, we emphasize only the secure variations, i.e., the significant changes observed by all instruments (EPIC+RGS). 

Our analysis is mainly based on the count rate lightcurves as they provide the most model-independent approach to the data. The EPIC lightcurves are composed of equivalent on-axis, full-PSF count rates, so that any offset or bad pixel problem was corrected to provide comparable lightcurves for all exposures (this possibility is not available for RGS). With this correction, all three EPIC detectors are considered to be stable over the spacecraft's lifetime by the XMM calibration team. Therefore, the flux found from spectral fitting and the count rate derived in the same energy band are always related by a constant factor for a source with a constant spectral shape (which is the case of \zp, see Paper I): using one or the other will yield exactly the same result, and fluxed lightcurves will thus not convey more information than those made from count rates. It should also be noted that fitting the spectrum of \zp\ is actually not an easy task (see Paper I), since we are reaching the limits of the instruments and atomic data precision. Within these limits and within the errors, the spectral shape of \zp\ is not changing from one observation to the next (see Paper I), but one could wonder whether this is true within an exposure. However, this would imply fitting spectra extracted in short time bins, hence very noisy. Such fitting would be highly unreliable, unstable and it may even cause spurious changes in the spectral parameters, due to the complex spectral shape of \zp, the many local minima, and the uncertainties in atomic data. These spurious changes will thus blur the actual source variations, rendering the task of studying them impossible in practice. Working with count rates avoid these problems, yielding clear and direct information on a given energy band, if not on a given spectral parameter other than flux.  

\subsection{EPIC lightcurve production}

Lightcurves of the source and background were extracted using the SAS task {\it epiclccorr} for four time bins (200s, 500s, 1ks, 5ks) and in seven energy bands: total (0.3-4. keV), Bergh\"ofer's band (0.9-2. keV), soft (0.3-0.6 keV), medium (0.6-1.2 keV), hard (1.2-4. keV), very hard (4.-10. keV) and grand total (0.3-10. keV). The background and source regions are the same as for the spectral analysis (see Paper I).

\begin{figure}
\includegraphics[width=8cm,bb= 30 200 560 640, clip]{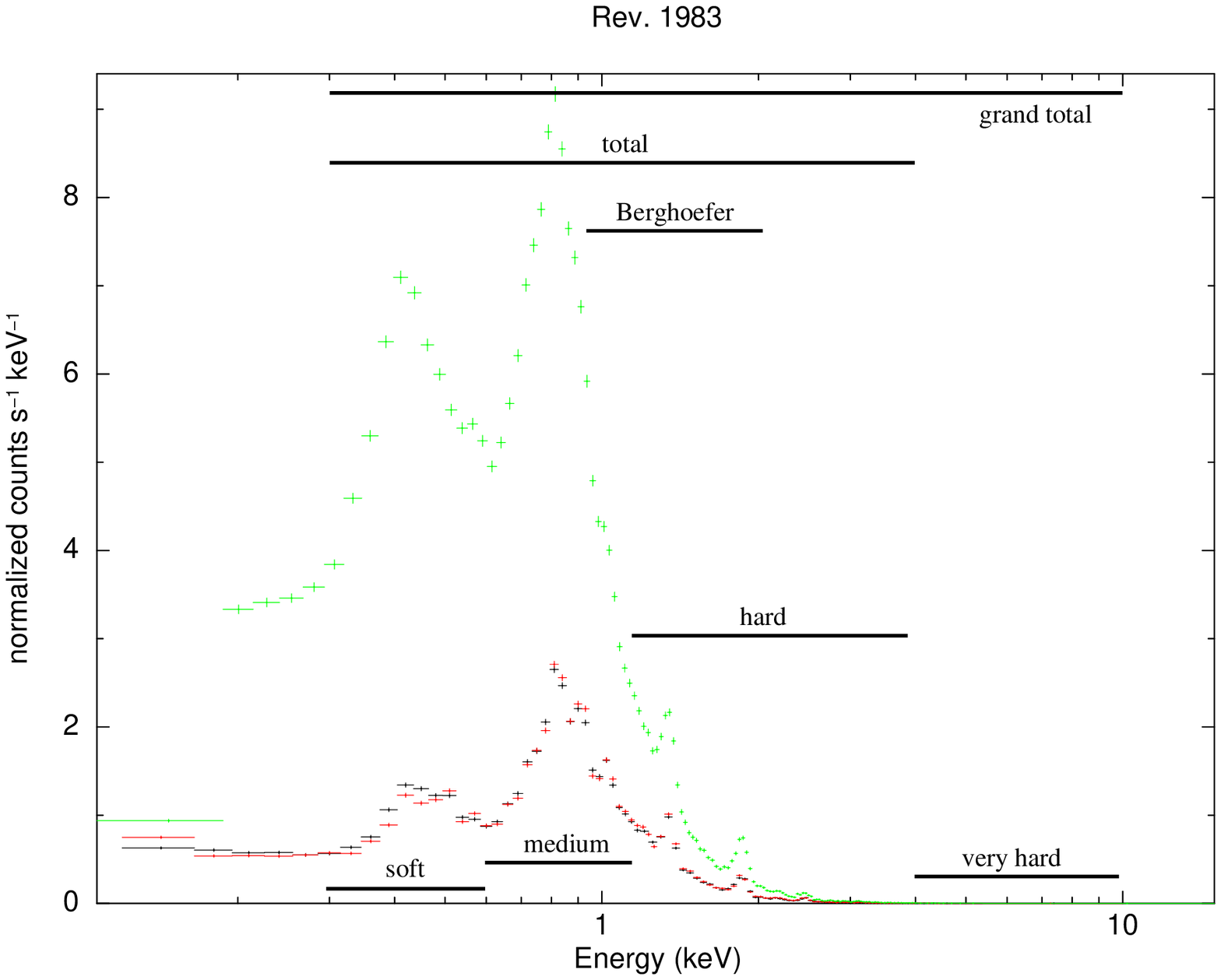}
\includegraphics[width=8cm,bb= 30 200 560 640, clip]{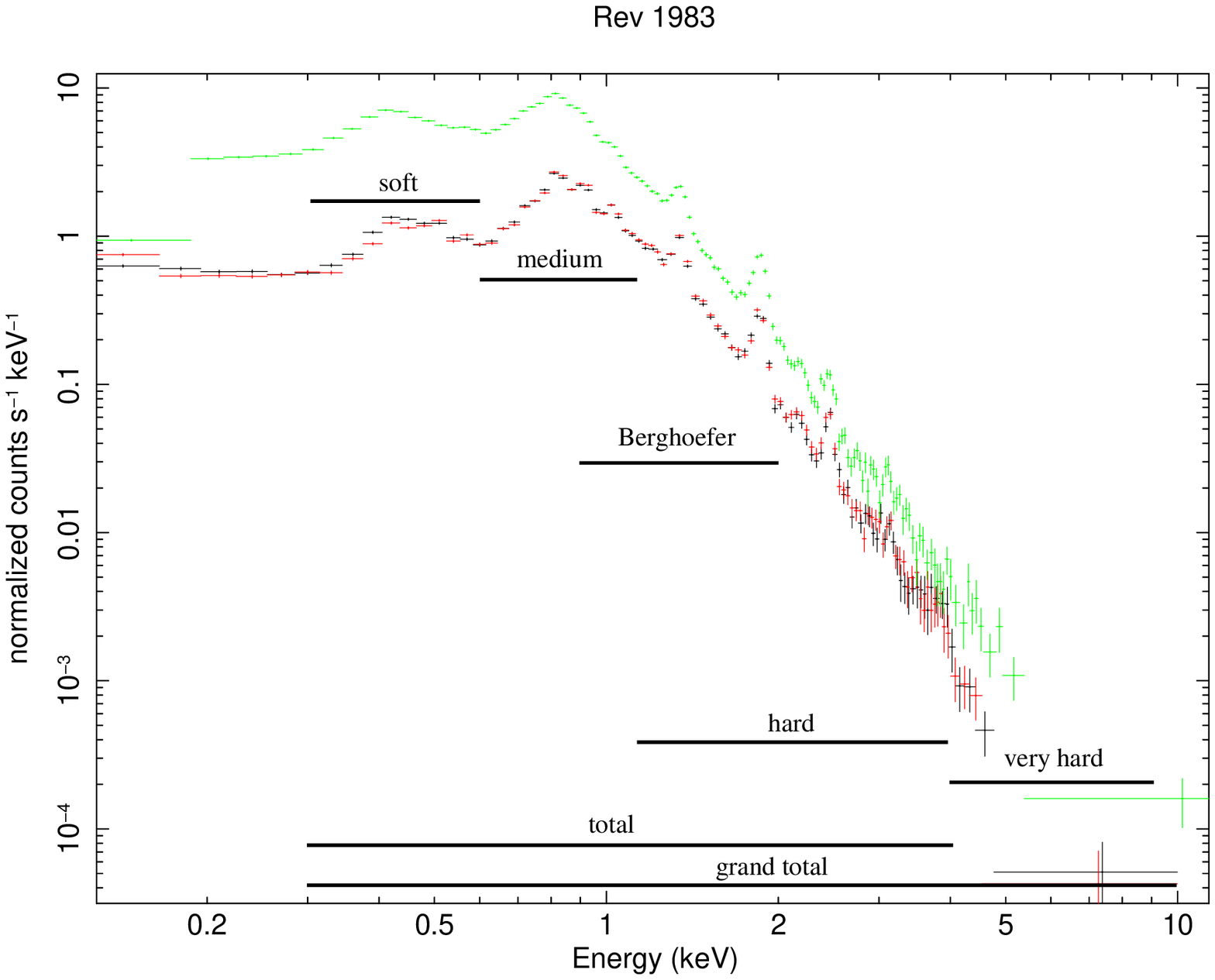}
\caption{Position of the EPIC bands compared to the spectra from Rev. 1983 (top  green points: pn, bottom black and red points: MOS).}
\label{epicband}
\end{figure}

The choice of the energy bands results from a compromise, taking into account the appearance of \zp\ spectrum, count rates in each band, and previous results. Indeed, one has to test bands for which results were reported before. As explained in Sect. 2, a periodicity was found using $ROSAT$ in the 0.9-2\,keV band. This is why we define a ``Bergh\"ofer's band'', to be able to confirm (or not) the presence of this signal in our much higher quality data. Another obvious choice is the total band, which enables us to use all available counts, and therefore get the smallest error, performing the most sensitive test of the overall level of the X-ray emission. Additional narrower bands (soft, medium, hard) are defined by taking into account the general spectral energy distribution (Fig. \ref{epicband}). The soft and medium bands probe the brighter part of the spectrum (with count rates above $\sim$0.6\,cts\,s$^{-1}$\,keV$^{-1}$), while the hard band enables us to investigate the tail of the spectrum. EPIC spectra showed a dim region around 0.6 keV, hence the choice of this energy to split soft and medium energy bands. Count rates are rather similar for these ``narrow'' bands: 0.8, 0.4, 1, and 0.4 cts\,s$^{-1}$ for MOS (resp. 2, 2, 3.5, and 1 cts\,s$^{-1}$ for pn) in the Bergh\"ofer, soft, medium, and hard bands, respectively. The spectra appear very noisy above 4 keV, hence the choice of that energy as an upper limit for the main energy bands. Nevertheless, a (double) check was made as we want to ensure that including all data with $>$4 keV does not change the conclusions. Indeed, the grand total and total bands provide indistinguishable results. The very hard band would reveal the presence of very high energy phenomena (transient or not). For \zp, however, this band always presents a very low count rate, of the order of $\sim(10\pm3)\times 10^{-4}$\,cts\,s$^{-1}$ for MOS (1 or 2), i.e. 3 orders of magnitude below the total band count rate, and similar to the background count rate in the same band for pn. This means that no bright emission at very high energies is present in this star, and we will therefore focus the analysis of \zp\ on the five main energy bands. 

The choice of the time bins was also made by taking several factors into consideration. First, we considered the problem of sampling different timescales. As one wished to study \zp\ on as many timescales as possible, both short bins and longer bins are highly desirable. However, the upper limit on the time bin length is dictated by the length of the individual exposures (13ks for the shortest ones) since at least a few bins are required for a meaningful analysis within the exposure itself. Second, we looked at the impact of errors. Our goal is to detect the most subtle variations of \zp. To this aim, we need to combine the smallest possible errors. 
Considering the count rates for the different bands and instruments (see above), we can reach a 5\% error on MOS data in 200s for the total band, in 500s for the medium and Bergh\"ofer bands, in 1ks for the soft and hard bands; a 3\% error on pn data in 200s for the total band, in 500s for the soft, medium and Bergh\"ofer bands, in 1ks for the hard band. A time bin of 5ks has been added to get 1\% or less error on the total band and enhance the detection of longer-term changes within one exposure. 

\begin{figure}
\includegraphics[width=8cm]{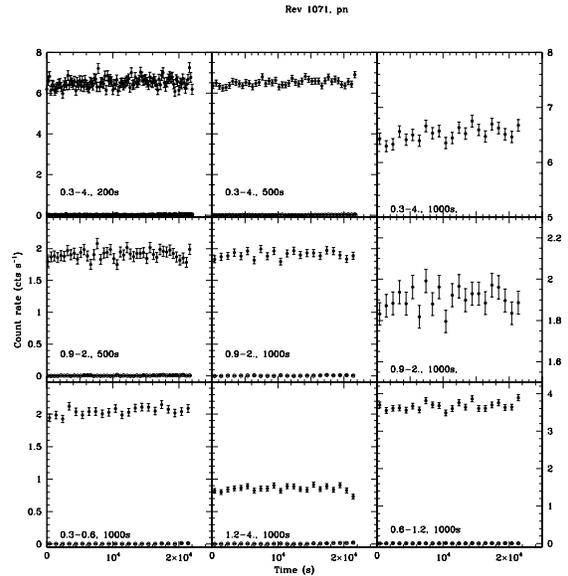}
\caption{ EPIC data registered for Rev. 1071 by the EPIC-pn instrument, for different bands and time bins. The background-corrected lightcurves of the source appear at the top of each panel, while the background lightcurves appear at the bottom. The x-axis represents elapsed time (in seconds) since the beginning of the observation.}
\label{example}
\end{figure}

To check the influence of the background (which is much fainter than the source in all observations), three sets of lightcurves were produced and analyzed individually: the raw source+background lightcurves, the background-subtracted lightcurves of the source and the lightcurves of the sole background region. The results found for the raw and background-subtracted lightcurves of the source are indistinguishable. An example of a lightcurve set is given in Fig. \ref{example}.

Finally, two remarks must be made. To avoid very large errors and bad estimates of the count rates, we discarded bins displaying effective exposure time $<$50\% of the time bin length. Our previous experience with \xmm\ has shown us that including such bins degrades the results. It should also be noted that the SAS task {\it epiclccorr} provides equivalent on-axis, full PSF count rates, so that problems such as the presence of an offset or of bad pixels were corrected to provide comparable lightcurves for all exposures.

\subsection{RGS lightcurve production}

The task {\it rgslccorr} yielded raw source+ background lightcurves, background-corrected lightcurves for the source and background lightcurves for ten wavelength bins: total (6-30\AA, equivalent to 0.4-2 keV), Bergh\"ofer's band (6-14\AA, equivalent to 0.9-2 keV), soft (20-30\AA, equivalent to 0.4-0.6 keV), medium (14-20\AA, equivalent to 0.6-0.9 keV), N{\sc vi} (28-30\AA), N{\sc vii} (24.3-25.5\AA), O{\sc vii} (21.3-22\AA), O{\sc viii} (18.75-19.2\AA), Ne{\sc ix} (13-14\AA), and Ne{\sc x} (11.8-12.5\AA). They were calculated for each instrument (with both orders combined) and for the whole RGS (with both instruments and both orders combined). Note that the wavelength of the O{\sc vii} line is at a gap of RGS2 while Ne{\sc x} is in a gap of RGS1 (order 1).

\begin{figure}
\includegraphics[width=8cm]{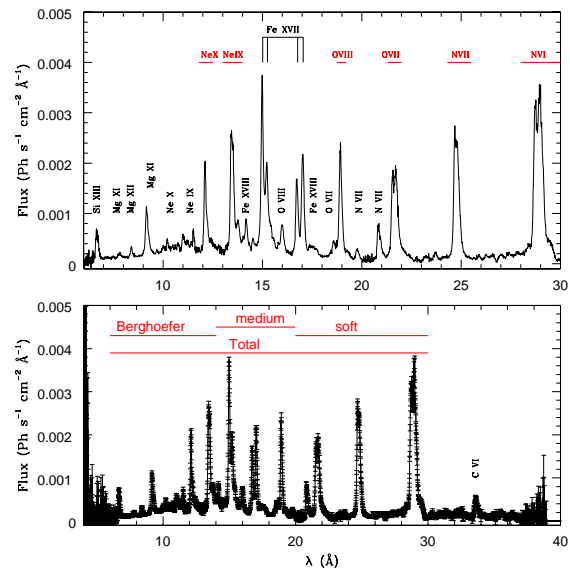}
\caption{Position of the RGS bands compared to the combined high-resolution spectrum (bottom: full with error bars, top: zoom on the 10-30\AA\ region).}
\label{rgsband}
\end{figure}

As for EPIC, there were obvious choices, such as the total band or bands covering the strongest lines, but also choices dictated by previous reports (Bergh\"ofer's band), and compromises (soft and medium bands - the soft band is similar to its ``twin'', the EPIC soft band, while the medium band appears in between soft and Bergh\"ofer's bands).  The limits of the wavelength intervals were carefully chosen taking into account the global aspect of the high-resolution spectra (Fig. \ref{rgsband}). Limits of the bands fall in regions as free of lines as possible; the overall lower and upper limits are dictated by the increasing noise towards short and long wavelengths. It may be noted that count rates for the soft, medium and Bergh\"ofer bands are similar (about 0.5\,cts\,s$^{-1}$). The bands covering lines were chosen to enclose the brightest isolated lines without enclosing too much of the neighbouring continuum.

Again, as for EPIC, the choice of time bins represents a compromise between getting a sufficient number of counts to perform an analysis with a few percent relative error only and having enough bins within each exposure (of length 30 ks for the shortest ones). Bins of 500s and 2ks yield errors in the total band, for each RGS, of  7\% and 3.5\%, respectively; bins of 2ks and 5ks give errors in the soft, medium and Bergh\"ofer bands of  7\% and 3.5\%, respectively; bins of 5ks and 10ks ensure errors of 10\% and 5\%, respectively, for bands linked to specific lines.

Note that, contrary to {\it epiclccorr}, the SAS task {\it rgslccorr} does not correct for lower recorded fluxes if the source appears off-axis. While this does not appear to be a problem for small offsets ($<$1.5'), the count rate of \zp\ during Rev. 0731 (where the offset reaches nearly 6') is clearly reduced, by about 25\% in the total band. In addition, the count rate from Rev. 0091 at highest energies (shortest wavelengths) appears too high by about 20\%. The origin of this problem is unknown (poor calibration at the earliest times of \xmm\ operations or true brightening? - there is no EPIC data to confirm what happened). To avoid any interpretation problems, both datasets were discarded from global analyses (see Sect. 4 and 5). 

It is also important to remember that {\it rgslccorr} performs a randomization of the time tags, i.e. all runs of the same task, with the same parameters and input files will produce slightly different results (the difference remaining within the error, of course). This is not the case of {\it epiclccorr} since a randomization is applied during the initial processing of the raw files (tasks {\it epproc, emproc}).

\section{Long-term variations (months to years)}

\subsection{Count rates}

To search for the long-term variations on timescales of months to years, we computed the average count rates for each exposure and checked their constancy using $\chi^2$ tests. We remind the reader of a few cautions. For RGS, the data points associated with Rev. 0731 are discarded from global analyses since the effective area changes due to the off-axis position of \zp\ are not taken into account in {\it rgslccorr}. The data from Rev. 0091 are also discarded since they yield suspiciously high count rates (see XMM user handbook\footnote{http://xmm.esac.esa.int/external/xmm\_user\_support/documentation/uhb/index.html} and Sect. 3). For EPIC, the pn data taken with the Medium filter are most probably still slightly affected by pile-up (see also Paper I). Indeed, they yield clearly erratic results: while Revs. 535, 538 and 542 provide very similar count rates for their Thick filter data, this is not the case for their Medium filter data.

\begin{figure*}[htb]
\includegraphics[width=8cm,bb=50 170 380 710, clip]{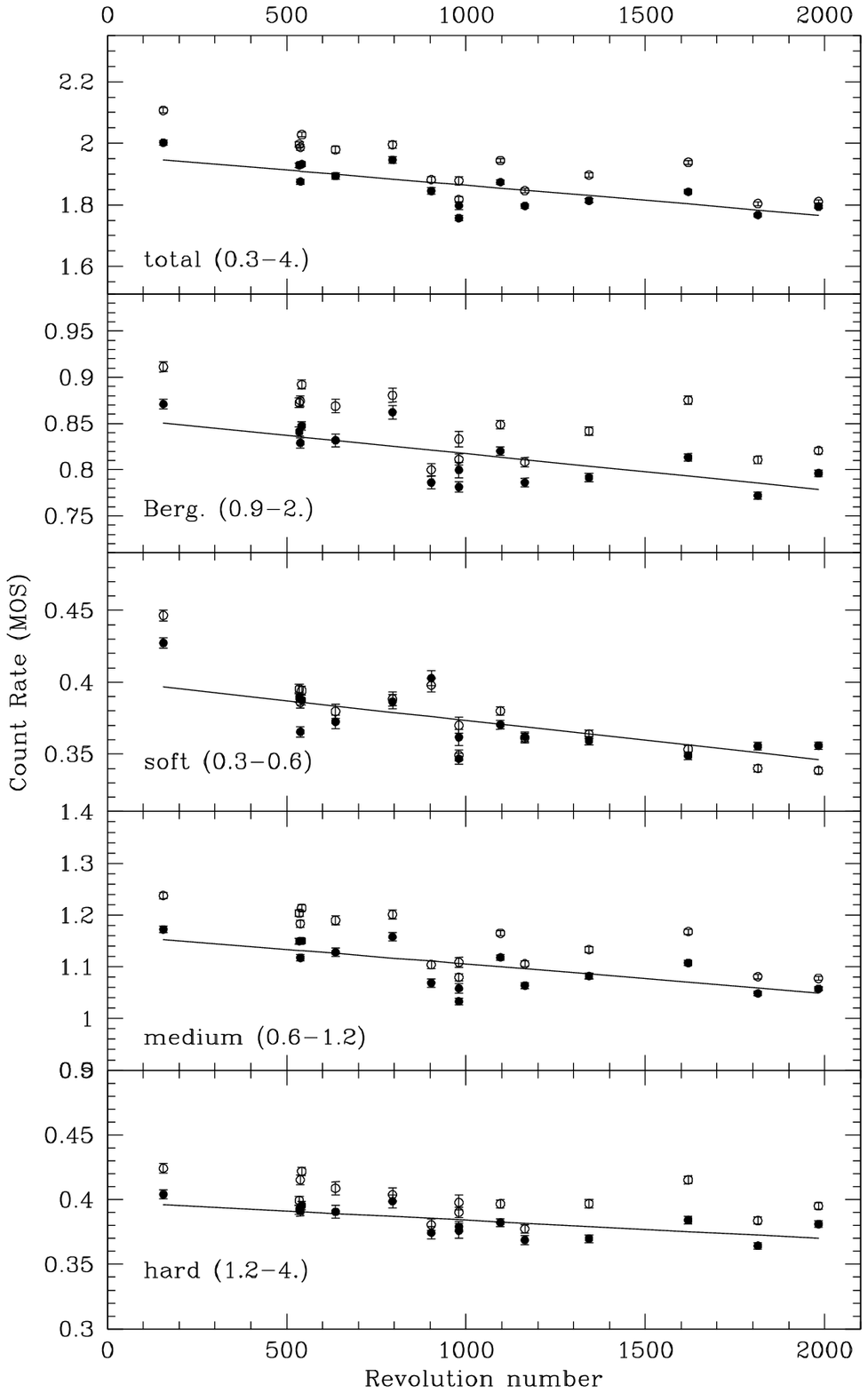}
\includegraphics[width=8cm,bb=50 170 380 710, clip]{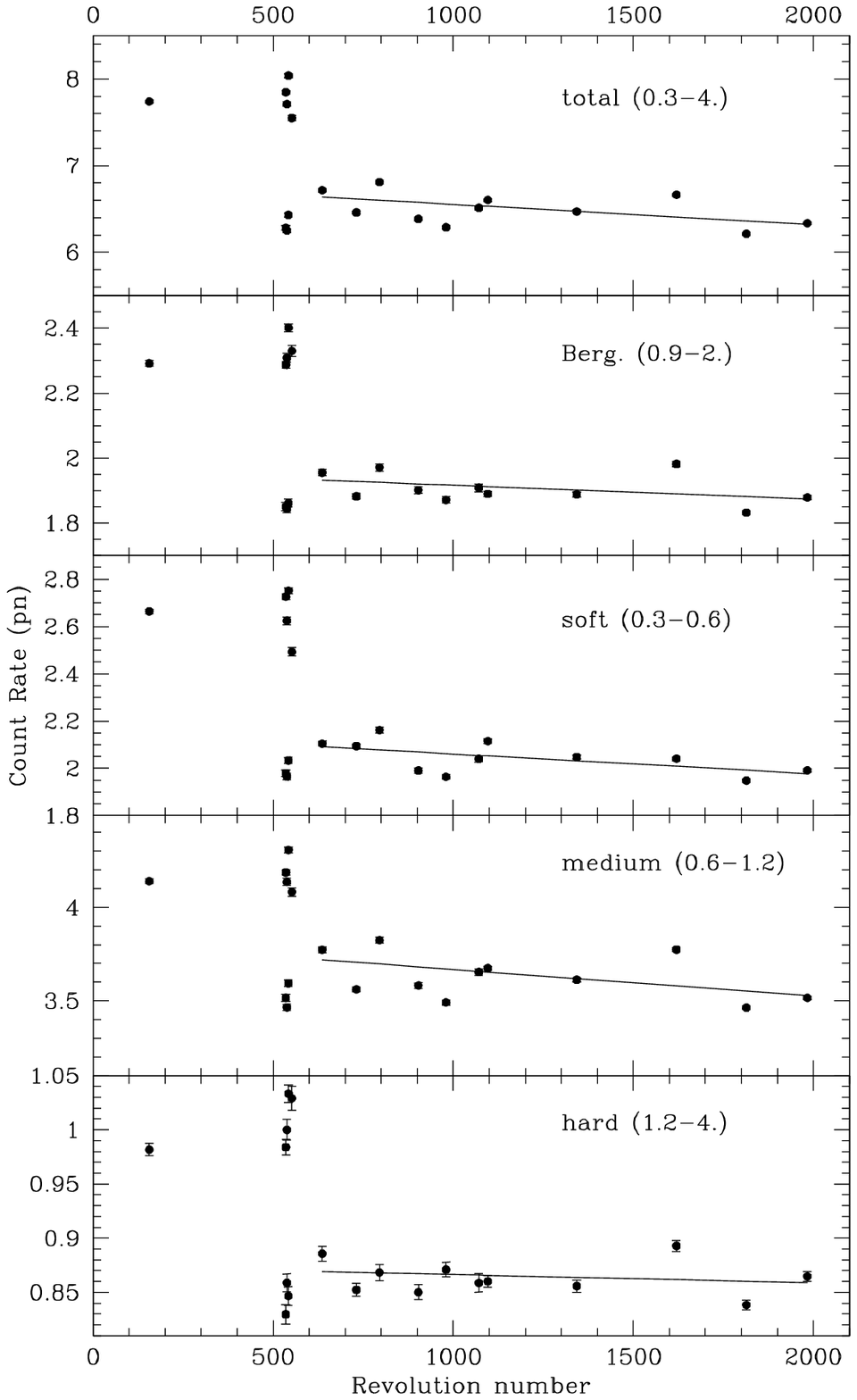}
\caption{Evolution of the average count rate with time. For MOS (left panel), filled and open circles correspond to MOS1 and 2, respectively. For pn (right panel), the 5 points at the beginning of the dataset appearing above others were taken with the Medium filter rather than the Thick one, and they are more affected by pile-up (see Paper I for details). In all panels, the best-fit linear trends (excluding the discrepant observations, see text) are drawn - for MOS, only that of MOS1 is shown for clarity.}
\label{totallc}
\end{figure*}

\begin{figure}
\includegraphics[width=8cm,bb=50 240 380 710, clip]{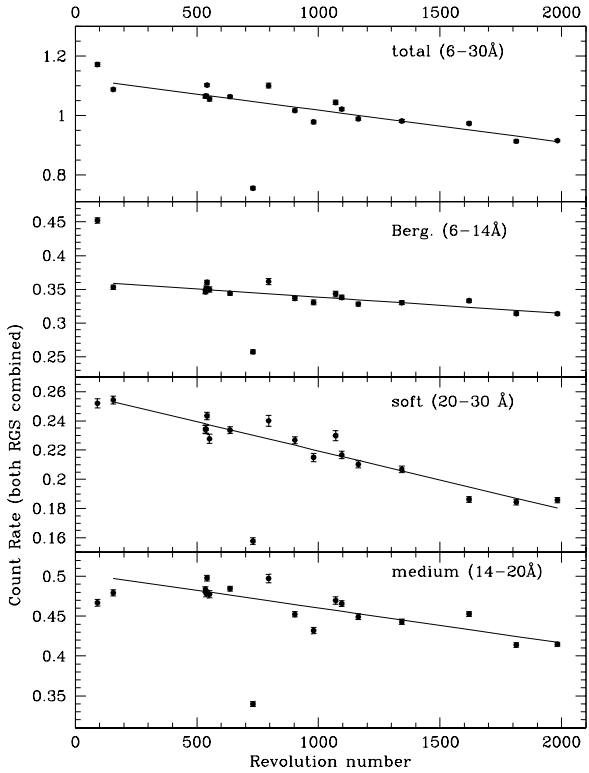}
\caption{Same as Fig. \ref{totallc} for the RGS data. The point which is systematically lower than others is associated with Rev. 0731, for which \zp\ was 6' off-axis (see text for details).}
\label{totallc2}
\end{figure}

Thereby considering only the best (i.e. most reliable) data, it clearly appears  that the count rates are not stable over timescales of years (Figs. \ref{totallc} and \ref{totallc2}). The lightcurves obtained by {\em each} detector decline. The rate of decline however differs among them. Over the $\sim$3800 days covered by the datasets, the EPIC-pn count rate decreased by about 6\% in the total, soft and medium energy bands, 2\% for the hard band and 4\% for Bergh\"ofer's band. The count rates of EPIC-MOS decreased by about 10\% in the total, medium and Bergh\"ofer's bands, 6\% in the hard band and 15\% in the soft band. The RGS count rates decreased by 18\% in the total band, 28\% in the soft band, 16\% in the medium band, and 12\% in Bergh\"ofer's band. These variations are  significant at the $<$1\% level\footnote{ A significance level of 1\% implies that there is only 1\% chance to get the observed deviation from the hypothesis by coincidence.} as the error bars on the average count rates are very small. 

The fitting of EPIC spectra also showed a small decrease in flux of a few percent (Paper I). 
Moreover, the supernova remnant 1E0102--72, observed independently from \zp\ (though also for calibration purposes), shows a decrease in its MOS count rate and flux similar to what is seen for \zp\ (M. Guainazzi, private communication). The exact origin of this systematic trend is not known (investigations are under way by the calibration teams), but such long-term effects are reminiscent of detector sensitivity ageing problems. Up to now, the pn and MOS are considered to be stable by the \xmm\ calibration team (M. Guainazzi, private communication)\footnote{See also Sect 1.3.2 and Fig. 1.14 of the EPIC calibration status which can be found on http://xmm.vilspa.esa.es/docs/documents/CAL-TN-0018.pdf as well as Sect. 5 and Fig. 5 of the XMM cross-calibration status available on http://xmm.vilspa.esa.es/docs/documents/CAL-TN-0052.ps.gz }, but there are apparently remaining imperfections in the long-term calibration. Since the decline of the registered count rates is currently not taken into account in the XMM calibration, we removed this trend by hand for further analyses and conclude that it is instrumental in nature.

Note that true variations of the source may be superimposed to these global decreasing trends. For example, \zp\ appears brighter during Rev. 1620 for both count rates and fluxes (derived from spectral fits, see Paper I). This is linked to the changes observed on intermediate timescales (see Sect. 5) - with trends on the course of days, the source may indeed appear brighter or fainter sometimes, depending on the observational sampling of these changes.

Finally, we also calculated the ratios between the RGS count rates associated to lines of the same elements - O\,{\sc vii}/O\,{\sc viii}, Ne\,{\sc ix}/Ne\,{\sc x} and N\,{\sc vi}/N\,{\sc vii}. Comparing globally the 16 secure RGS exposures (see above and Sect. 3), there is no obvious correlation between the different elements: the Ne\,{\sc ix}/Ne\,{\sc x} and N\,{\sc vi}/N\,{\sc vii} ratios do significantly vary, but only the N\,{\sc vi}/N\,{\sc vii} is clearly decreasing. The N lines constitute the most widely separated pair: as the long-term trend is instrumental in origin, such a calibration problem could affect some wavelengths more than others and a larger separation would then lead to larger differences in lines. The average values of these ratios, based only on count rates (i.e. not on dereddened fluxes) and excluding Revs. 0091 and 0731, are 1.44$\pm$0.01 for nitrogen (both RGS, both orders), 1.28$\pm$0.01 for oxygen (RGS1, both orders), and 1.86$\pm$0.01 for neon (RGS2, both orders) - the choice of instruments takes into account the fact that RGS2 could not record any information on O{\sc vii}, whereas the first order of RGS1 contains no data on Ne{\sc x} (see Sect. 3). 

\subsection{X-ray lines}

\begin{figure}
\includegraphics[width=9cm,bb= 120 170 550 700, clip]{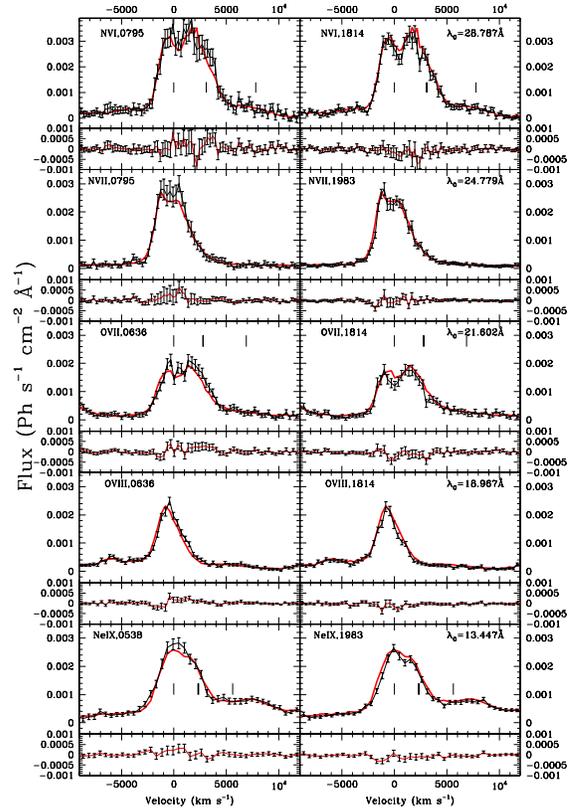}
\caption{Examples of ``large'' variations in X-ray lines for a few observations (identified by their revolution number in each panel). The individual spectrum, combining both instrument and both orders for a given revolution, is plotted with its errors and compared to the combined, full RGS spectrum at the same wavelength (red line) in each top panel. Their difference (in the sense individual spectrum minus combined spectrum) is given in each bottom panel.}
\label{linevar}
\end{figure}

High-resolution X-ray spectra of good quality can be derived for each exposure, and used to search for variations. The lines recorded for each observations can first be compared to the average spectrum obtained when combining all observations (Fig. \ref{linevar}). No large, significant variation is detected. Small changes in flux (of the order of one sigma) can be spotted from time to time. They are similar to those recorded at optical wavelengths \citep{eve98,rau10}, with double peaks where the blue to red ratio slightly changes. Their small amplitude and the low signal-to-noise of the data however prevent us from a detailed analysis: this must await the advent of more sensitive observatories.

\begin{figure}
\includegraphics[width=6cm]{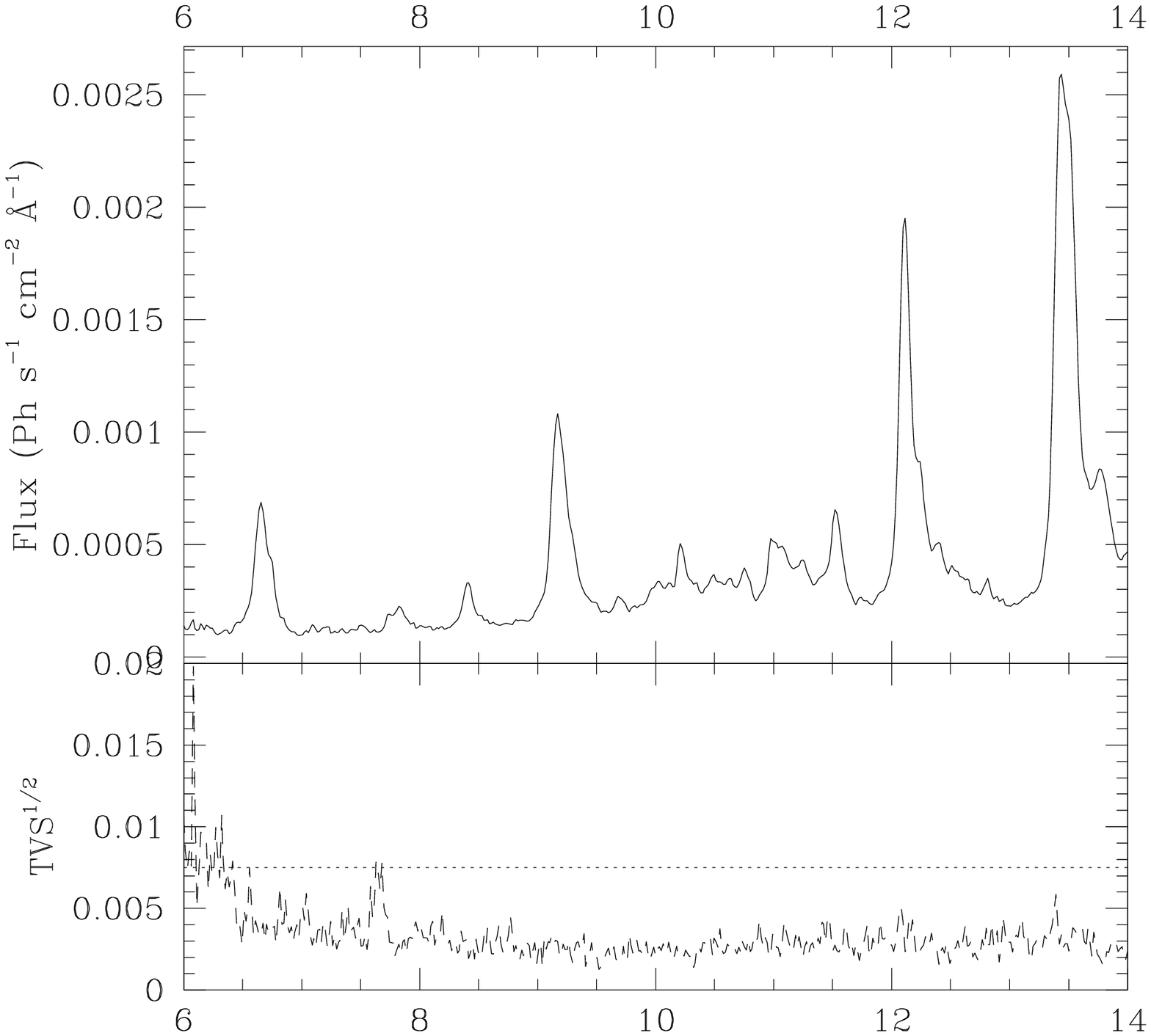}
\includegraphics[width=6cm]{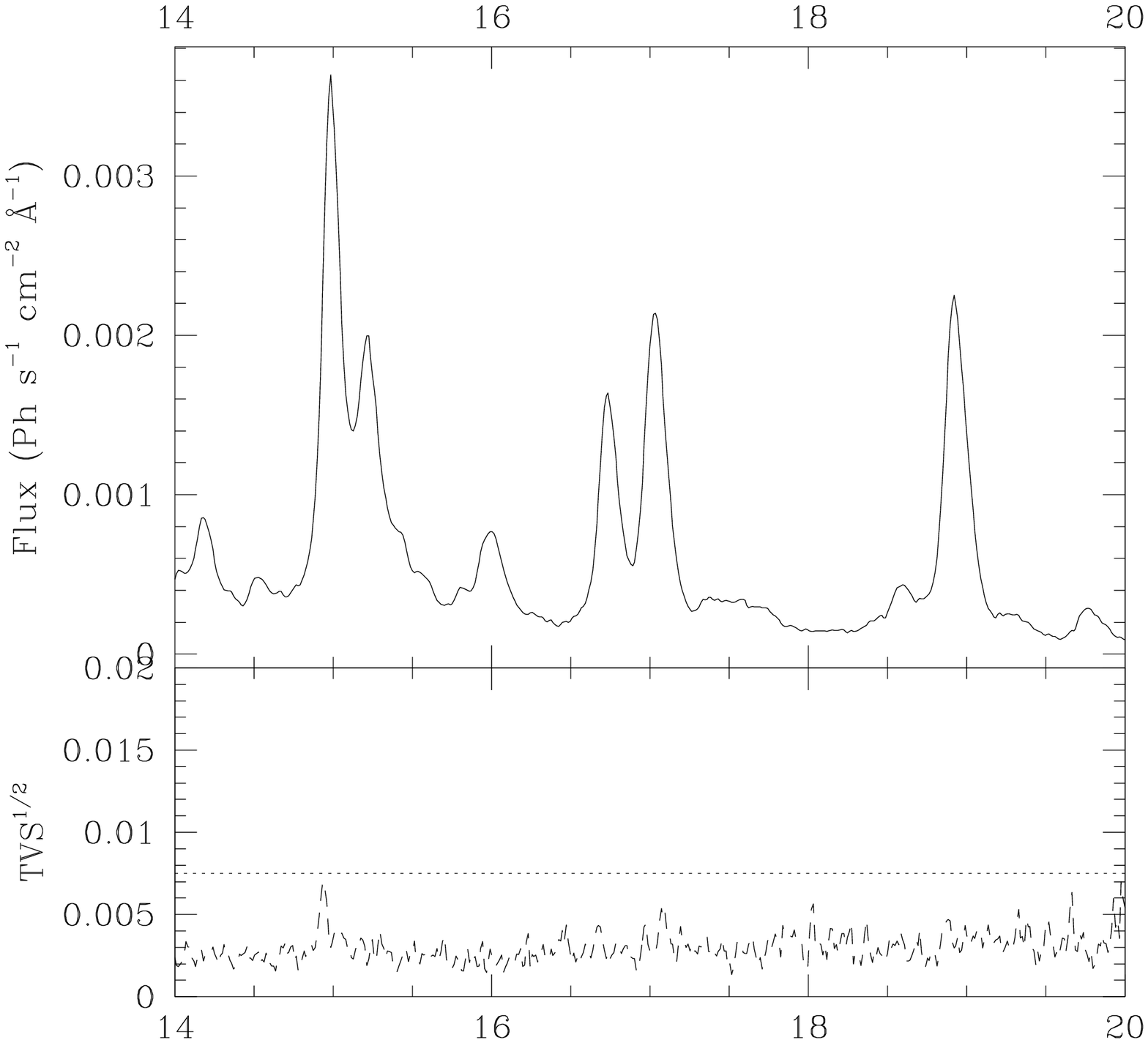}
\includegraphics[width=6cm]{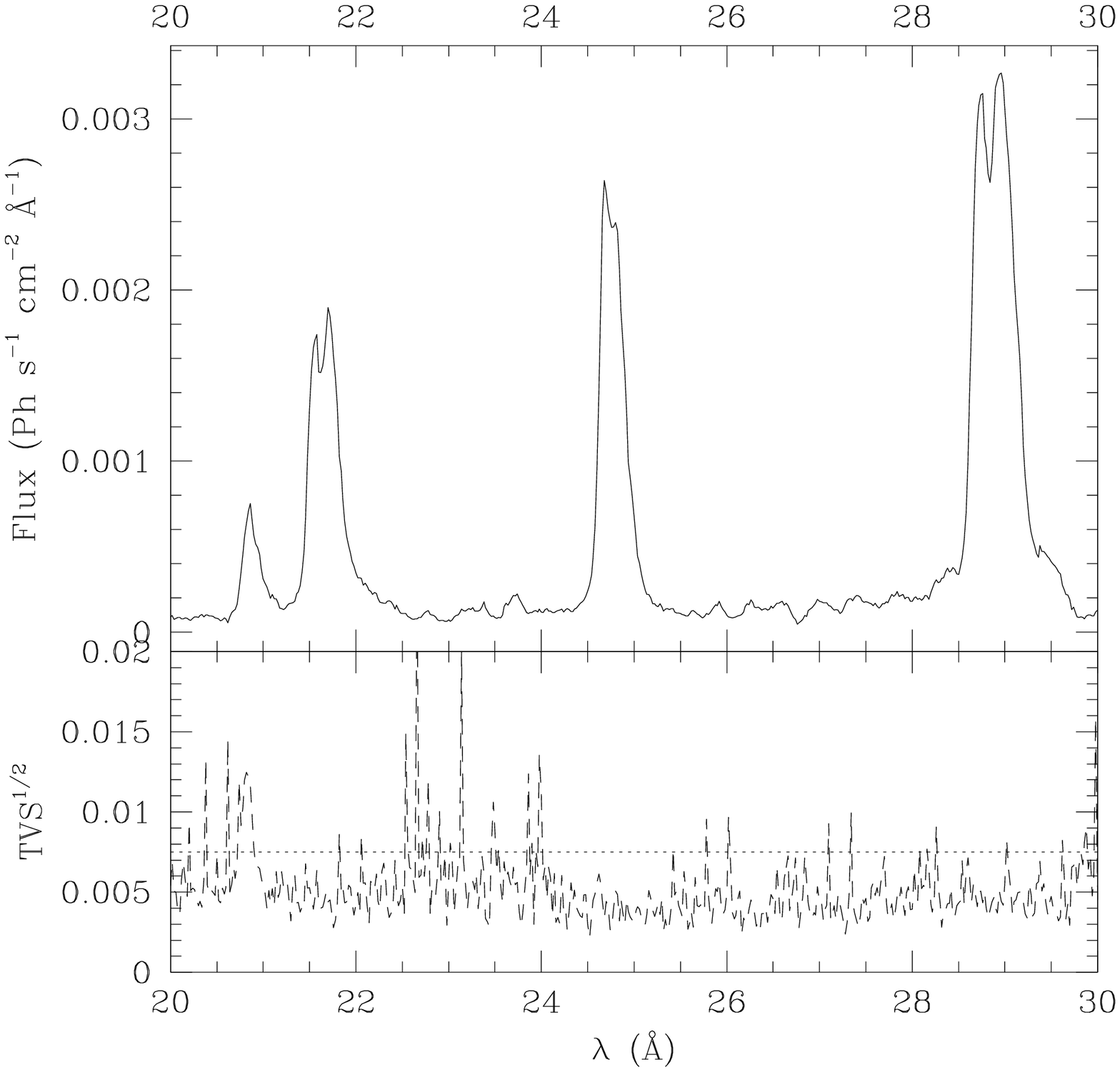}
\caption{The upper panels yield the mean spectrum of \zp, whilst the lower panels display the TVS. The dotted line corresponds to a significance level of 1\%.}
\label{tvs}
\end{figure}

To be more quantitative, a temporal variance spectrum analysis (TVS, \citealt{ful96}) was performed on the combined and fluxed RGS spectra (output of $rgsfluxer$).  The TVS computes the squared difference between individual spectra and the average one, taking the signal-to-noise ratios into account, which allows to detect statistically significant deviations from the average. Note that the spectra are calibrated both in flux and wavelength within $rgsfluxer$ (see Paper I for details), so that the pointing problems have no impact here. The relative weights for the different spectra in the TVS were chosen equal to the count rate errors on the total band. The TVS appears overall flat (implying constancy), with only few distinctive features (Fig. \ref{tvs}). For example, some peaks are found in the 20--24\AA\ region. This region is affected by gaps and missing CCDs, and there are therefore undefined values for wavelengths where no exposure exists. The exact position (in wavelength) where that occurs varies from one exposure to the next, notably because of pointing differences, and that results in apparent variability in the spectral set (hence peaks in the TVS). Increased noise at the shortest wavelengths (below 7\AA) and longest wavelengths (above 24\AA) is also producing a larger TVS, unrelated to intrinsic variability of the source (though the peak at 7.7\AA\ remains unexplained). Globally, the TVS thus indicates that the lines in the X-ray spectrum of \zp\ are not significantly varying from one exposure to the next. Very small changes are not excluded, however. Indeed, two small peaks of the TVS occur at 13.5 and 15\AA, i.e. at the wavelength of the strongest lines: this indicates that the line profile variations of \zp\ are just beyond the reach of current facilities, and they may thus be detected in the future with the better sensitivity of new X-ray observatories. 

\section{Intermediate-term variations (day to months)}

In this section, we investigate the data for the presence of variations with timescales of day to months. Within each exposure, the same set of tests was applied to each lightcurve. We first performed a $\chi^2$ test on all available individual lightcurves (i.e., 4 time bins and 7 energy bands for EPIC, 10 energy bins and 2 time bins for RGS, see Sect. 3) for several null hypotheses: constancy, linear variation, quadratic variation. We further compared the improvement of the $\chi^2$ when increasing the number of parameters in the model (e.g. linear trend vs constancy) thanks to F-tests (see Sect. 12.2.5 in \citealt{lin68}). A variability test using Bayesian blocks \citep[BBs, ][]{sca} was also performed, through the FTOOLS {\it battblocks}. It was made on the 200s full (i.e., including bins with $<$50\%  effective exposure time) EPIC lightcurves in the total energy band - those are the data with the largest signal-to-noise. For the most varying cases, a check was made by testing the event arrival times of the source, though this does not take into account the background, non-uniformities, or bad time intervals: results were similar to those derived from binned lightcurves. Unless otherwise stated the adopted critical significance level is 1\% throughout this paper. The individual results for each exposure are detailed in the Appendix, and we only summarize them below.

In general, the background is found to be variable. This is expected since many observations were affected by flares and some variations remain, despite the fact that the largest and narrowest ones have been cut out during the processing (see Paper I and Appendix for details).

For the source itself, however, only one thing is obvious: \zp\ does not vary much. For example, BBs are particularly useful to detect bursts, but there  are none in \zp, and a single block is found in most cases to be the best representation of the lightcurves. The absence of bursts is confirmed by the $\chi^2$ analyses. 

\begin{figure}[htb]
\includegraphics[width=9cm]{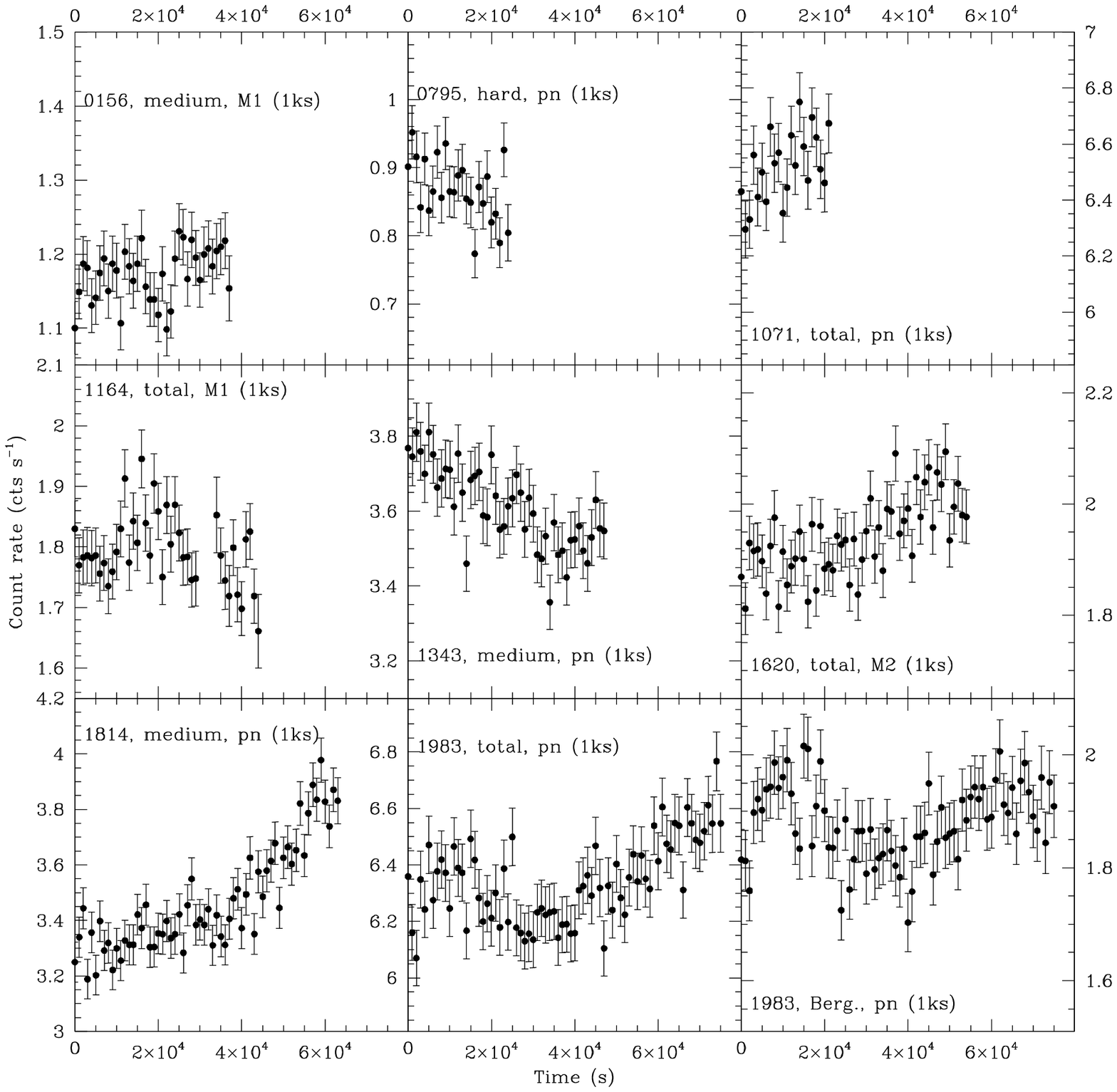}
\includegraphics[width=9cm]{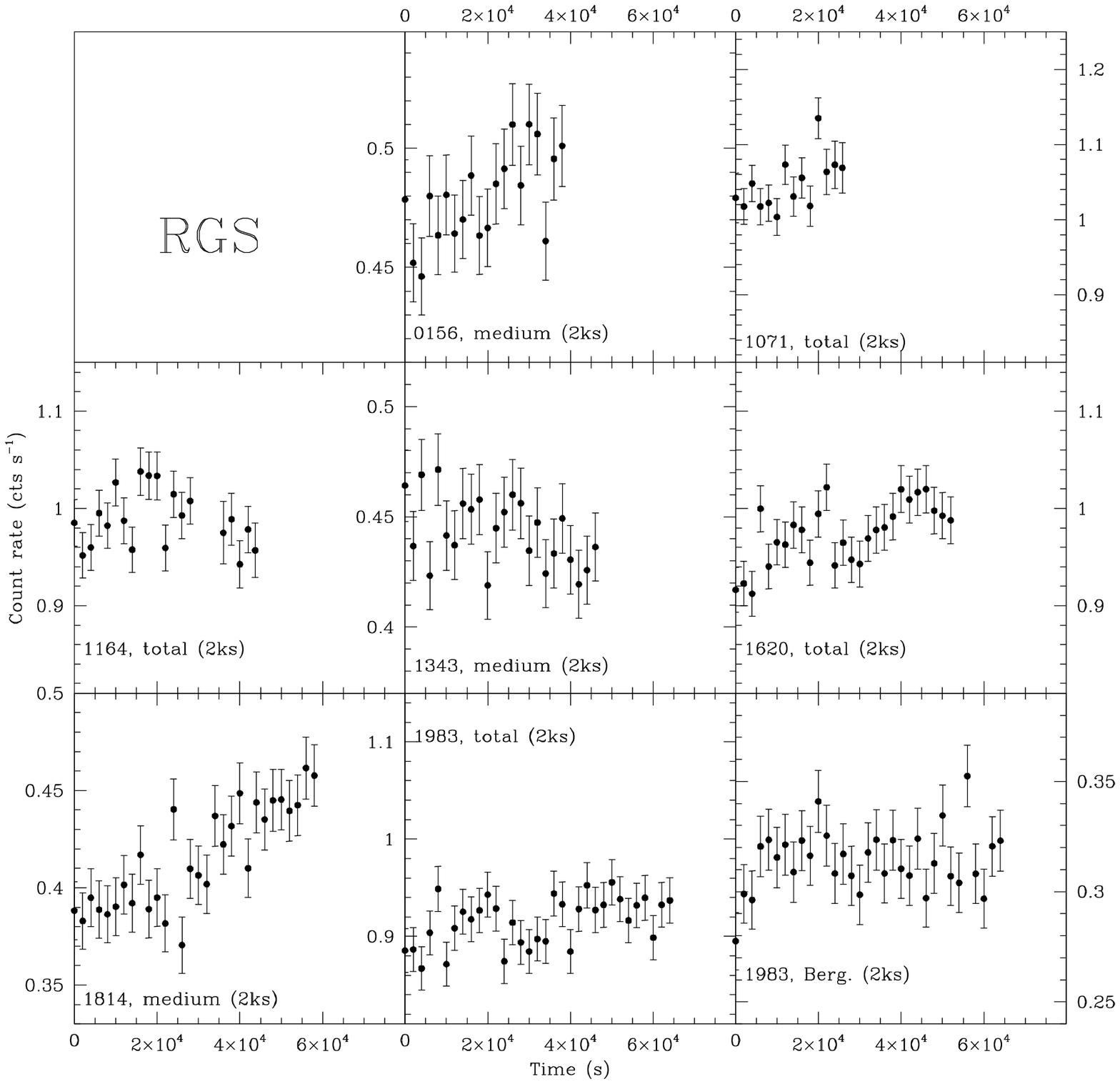}
\caption{A few examples of intermediate-term variations in lightcurves. The x-axis corresponds to elapsed time (in seconds) since the beginning of the observation.}
\label{revvariab}
\end{figure}

The error bars on each time bin are smaller for longer bins: small variations will thus be most easily detected for long bins. On the other hand, variations with timescales much smaller than the exposure duration will be  smoothed out when using long time bins. Therefore, if short-term variations were dominating the overall variability of \zp, the lowest variability level should be found for long time bins. What we find is the opposite trend: for both EPIC and RGS data, the longest time bins often yield the most variable lightcurves (see Fig. \ref{revvariab}) - though they rarely reach a significance level of 1\%. This indicates that trends with timescales similar or larger than the exposure duration are more common in \zp\ than very short-term events. 

\begin{figure}[htb]
\includegraphics[width=8cm]{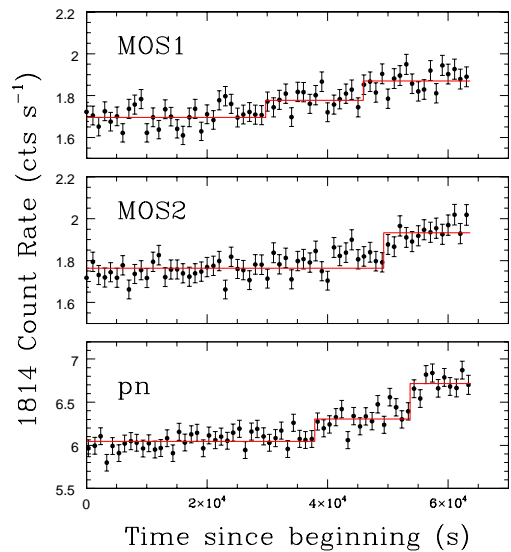}
\caption{ Evolution of the total count rate in Rev. 1814, with the Bayesian blocks superimposed. To ease the comparison, the corresponding lightcurves with 1ks time bins are shown. The x-axis corresponds to elapsed time (in seconds) since the beginning of the MOS1 observation.}
\label{bb}
\end{figure}

In fact, only 6 exposures show a significant non-constancy of their (mostly EPIC) lightcurves: Revs 0795, 1164, 1343, 1620, 1814, and 1983 (Fig. \ref{revvariab}). Moreover, in eight cases (Revs 0156, 1071, and the six previously quoted), modelling by a trend (linear and/or quadratic) yields a significant improvement of the $\chi^2$ over constancy, even if the significance levels attached to the individual $\chi^2$ (for the constancy hypothesis) are not $<$1\% in the additional two cases. It should be noted that 5 out of these datasets are amongst the longest exposures (see Paper I). Indeed, if we cut the data of Rev. 1983 to keep only the first 10, 20, 30, 40, 50 or 60ks, the $\chi^2$ (for the constancy hypothesis) never reaches a significance level $<$5\% and there is no significant improvement by fitting a line of non-zero slope rather than a constant! It thus seems that \zp\ appears variable each time one looks at it for a sufficiently long time. For the most varying cases, i.e. Revs. 1343, 1620, 1814, and 1983, a splitting of the lightcurves into two or three BBs is found (see e.g. Fig. \ref{bb}). Again, this favors the existence of longer-term shallow trends over that of shorter-term ``bursting'' events. 

The performed tests can also provide information on the energy at which variations occur, when they are detected. Usually, it is expected that the hard band would be the most often variable in O-stars, as highly variable phenomena such as colliding winds or magnetically-confined winds mostly produce hard X-rays. However, this seems not to be the case in \zp, as there is no coherent behaviour for this band - in some exposures, the hard band appears as the most variable (i.e. that with the largest dispersion or $\chi^2$), in others as the least variable. In contrast, the soft band has a more consistent behaviour: it is only detected as variable when all other bands are varying too; it can therefore be classified as the least often variable one. Within the observational limitations, the variations in \zp\ therefore appear as globally affecting its spectrum, rather than being particular to a specific (narrow) energy band.

\subsection{Fourier analysis}

The characteristic timescale of these intermediate-term variations is difficult to assess on individual exposures using $\chi^2$ tests or BBs. These first analyses can only conclude that the timescale is longer than the typical exposure length (i.e., a day or more). Looking further at the individual lightcurves, no obvious oscillation is detected by eye, despite the full coverage of the putative $ROSAT$ 17h-period by individual exposures and the large improvement in quality over the $ROSAT$ data. It thus seems that a sinusoidal variation of several hours and an amplitude of a few percents, such as that detected by \citet{ber96}, is transient, at best.

To give a more quantitative assessment of the variability timescale, we have performed period searches on global lightcurves (created by putting each individual-exposure lightcurve after one another, keeping their individual time tags), corrected for the long-term instrumental trend described in Sect. 4.1. We used the algorithms of \citet[see also correction by \citealt{gos01}]{hmm}. As in previous sections, we considered only the best datasets (see Paper I), i.e. the 15 MOS exposures and the 10 pn exposures taken in small window + Thick filter mode, as well as the 16 RGS exposures (i.e. excluding Revs. 0091 and 0731). The small time bins were favored (200s for EPIC, 500s for the total RGS band, and 2ks for the other RGS bands) since the Fourier-period-search algorithms are more affected by a small number of points than by large individual errors. Note that a more general period search technique which attempts to fit a period simultaneously with its harmonics does not yield additional insights. 

\begin{figure*}
\includegraphics[width=7cm,bb=40 170 380 710, clip]{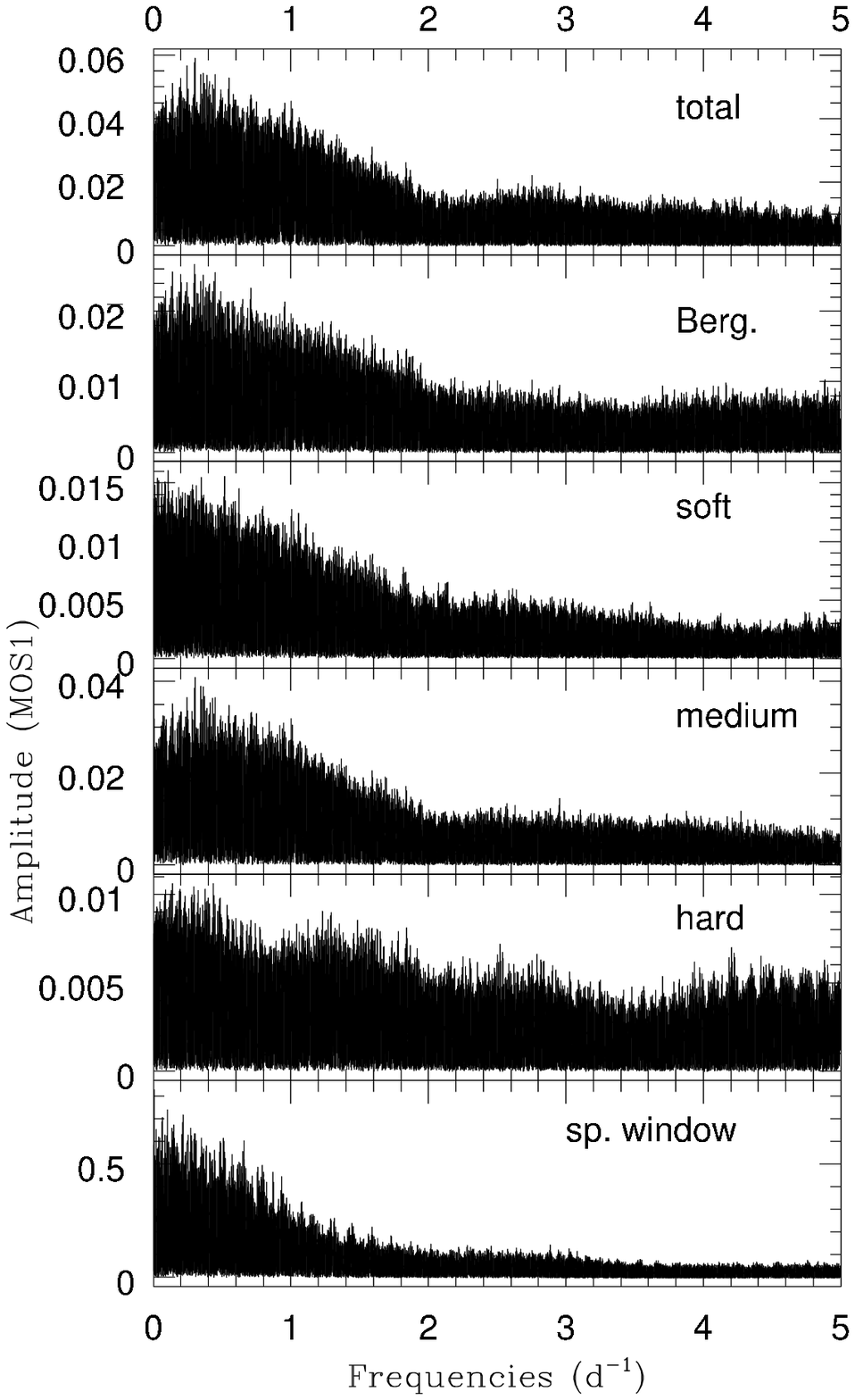}
\includegraphics[width=7cm,bb=40 170 380 710, clip]{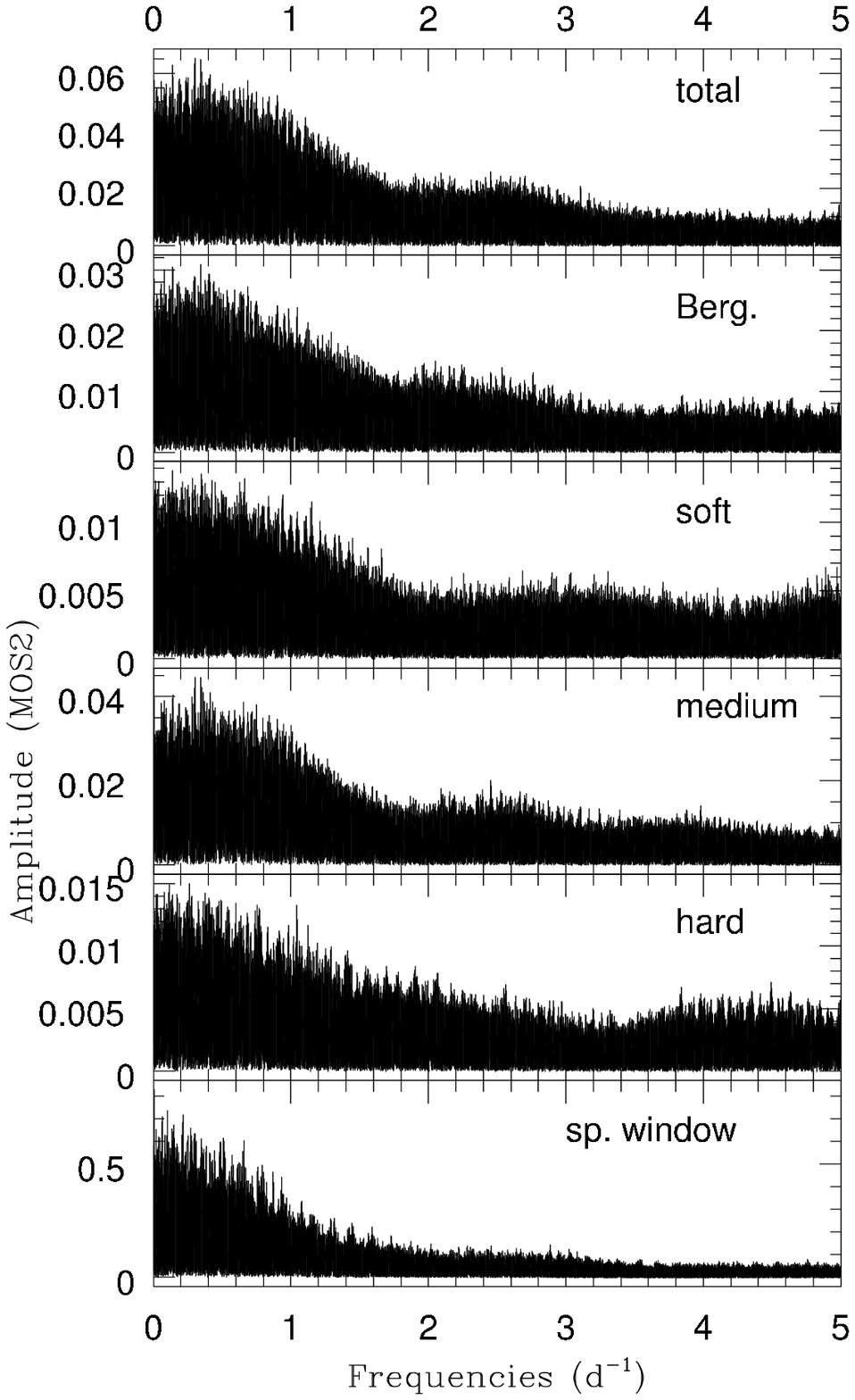}
\includegraphics[width=7cm,bb=40 170 380 710, clip]{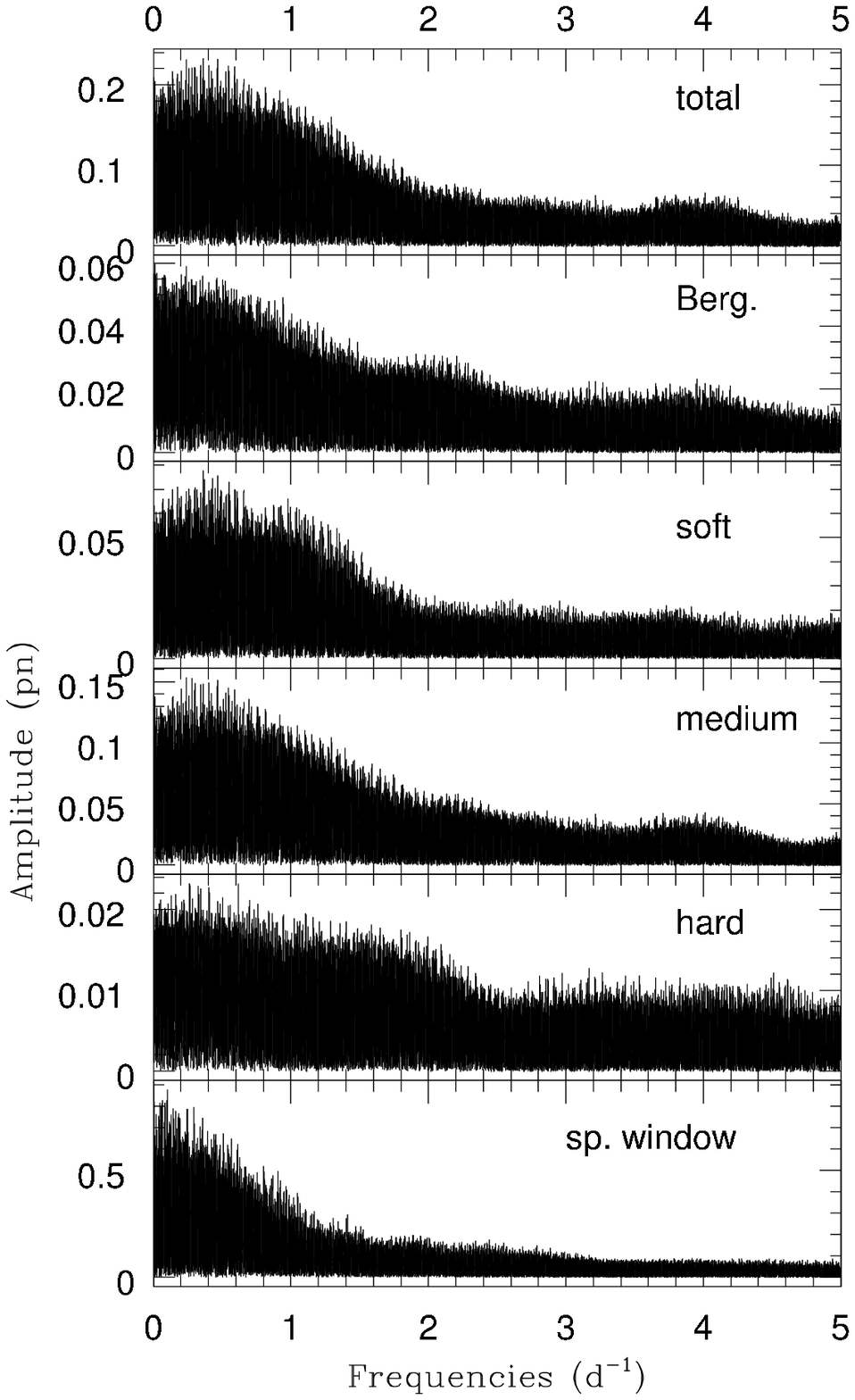}
\includegraphics[width=7cm,bb=40 240 380 710, clip]{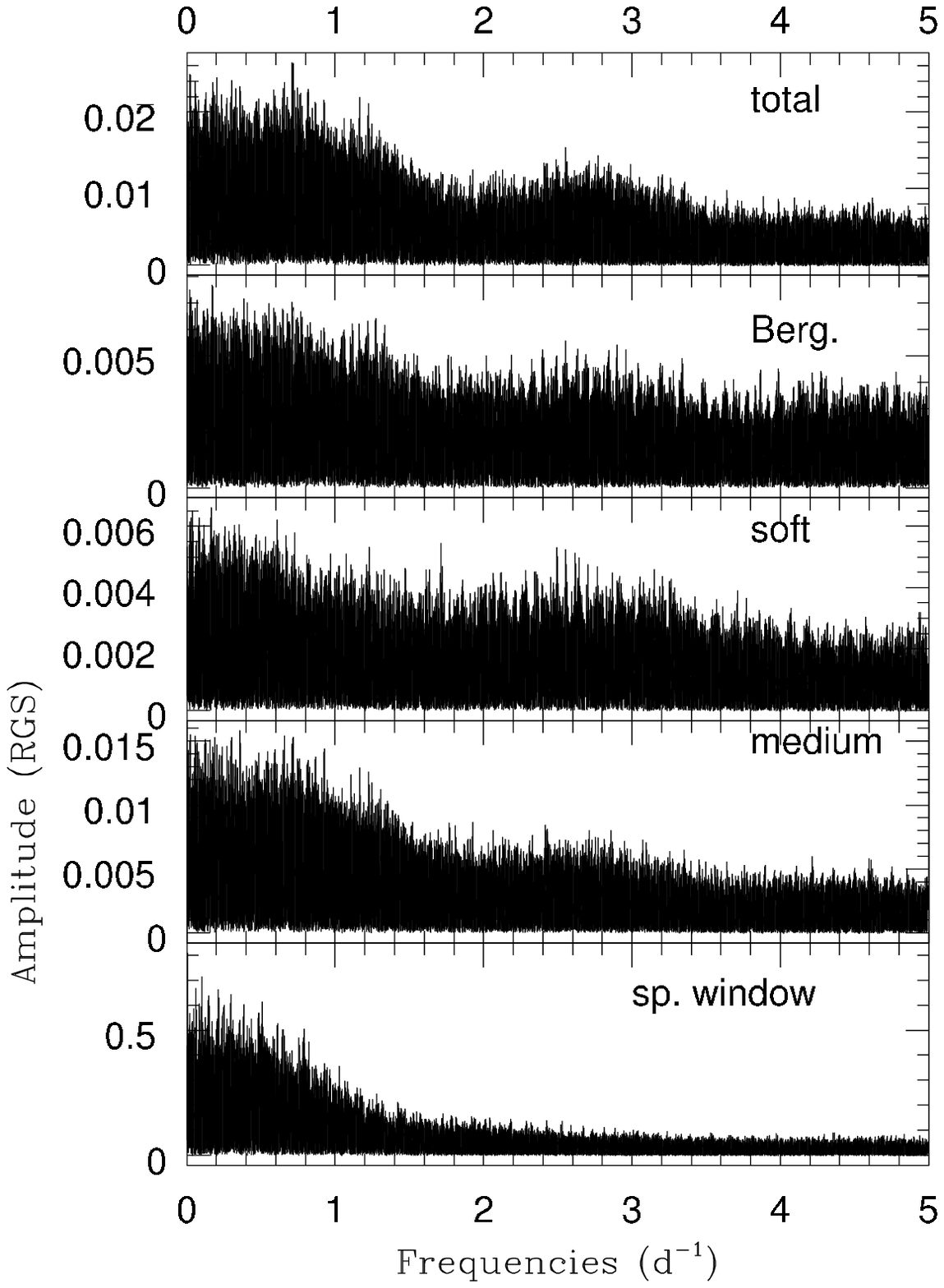}
\caption{Periodograms in different energy bands for the EPIC instruments and the combined RGS data.}
\label{fourier}
\end{figure*}

The resulting periodograms are shown in Fig. \ref{fourier}, along with the spectral window (i.e. the aliasing structure appearing naturally because of the temporal sampling). The periodograms consist of narrow peaks (since the dataset covers more than 10 years) tightly disposed in much broader features (because there are only a few exposures, of max.\ 70ks, within this 10-year timescale), which hamper an accurate determination of any period. However, two features are apparent for all bands except the hard one: (1) the highest peak occurs at about 0.3--0.4\,d$^{-1}$, (2) the right wing of this peak decreases much more slowly than that of the spectral window - some additional power is thus present around 0.7--1.3\,d$^{-1}$.

To assess the significance of these signals, we performed a formal iterative decomposition in frequencies - we here illustrate only the analysis for the pn total bandpass. A first extracted frequency at $\nu_1 \, = \, 0.365 \, \mathrm{d^{-1}}$ (period $\mathrm{P} \, = \, 2.74 \mathrm{d}$, semi-amplitude $\mathrm{a} \, = \, 0.25 \, \mathrm{cts/s}$) is significant against the null-hypothesis of white noise as deduced from ad hoc simulations. The following frequencies are $\nu_2 \sim \, 0.89 \, \mathrm{d^{-1}}$ ($\mathrm{P} \, = \, 1.12 \mathrm{d}$, $\mathrm{a} \, = \, 0.12 \, \mathrm{cts/s}$) and $\nu_3 \sim \, 0.4 \, \mathrm{d^{-1}}$ ($\mathrm{P} \sim \, 2.5 \mathrm{d}$, $\mathrm{a} \, = \, 0.08 \, \mathrm{cts/s}$ ; the last values are rather uncertain because they depend on the selected binning). These peaks in the Fourier periodogram are still  characterized by a significance level lower than 0.001. The white noise hypothesis is thus formally rejected. 

However, these frequencies do not correspond to the true content of the signal of the star. For example, when phased with the 0.365\,d$^{-1}$ frequency, individual lightcurves appear one after another, with no common phase interval. As explained above, some individual runs exhibit shallow trends. The time spanned by these runs combined with their scarcity over the covered decade  has an unfortunate consequence: the Fourier transform is always able to find a few frequencies that are combining the various trends in a very constructive way. However the plethoric presence of gaps in the time series and the concomitant large number of degrees of freedom is responsible for the positive combination more than the possible coherency of the signal. As a test, we detrended the individual runs with a linear function before merging them. We computed the periodogram for the new time series. All the previously reported candidate frequencies disappear from the list of frequencies. The dominant one, for the 200\,s binning, is $\nu_4 \, = \, 1.42 \, \mathrm{d^{-1}}$ ($\mathrm{P} \, = \, 0.70 \mathrm{d}$,  $\mathrm{a} \, = \, 0.054 \, \mathrm{cts/s}$) which corresponds to a peak whose significance level is around 0.001. If we detrend the X-ray lightcurves over the individual runs with a second degree polynomial, we obtain a time series that exhibits no outstanding peak and is thus in perfect agreement with the  white noise hypothesis. Therefore, although the white noise hypothesis is rejected, the alternative hypothesis to adopt remains unsure. The star exhibits weak although significant variations at a time scale between 0.7 day and several days but the sampling does not allow us to conclude to the existence of a coherent, fully deterministic signal above Poisson noise. The signal could be dominantly stochastic with some coherency at short timescale (some kind of red noise\footnote{ While white noise has, on average, the same amplitude whatever the frequency, red noise displays stronger amplitudes at low frequencies}). It is also possible that these changes are transient: they would then not appear as a strictly periodic signal in the global lightcurve. Another possibility is that any signal could be not strictly stationary, in phase or frequency, from one observing run to the next, as often seen for changes of the optical spectra in Oef stars \citep[ - it must be recalled that our target, \zp, belongs to this spectral type category!]{rau03}. 

In summary, it is certain that \zp\ displays variations with relative amplitudes up to 10--15\% on the timescales of days - not hours (i.e., there is no trace of the signal attributed to non-radial pulsations, see \citealt{rei96}). However, our dataset suffers from the misfit between the possible frequencies and the actual sampling. Therefore, despite its exceptional quality, it is still insufficient to detect confidently evidence for the presence of a coherent signal with daily timescales. 
A modulation of the X-ray flux with such daily timescale was reported for the O-type dwarf $\zeta$\,Oph, and potentially associated with the corotating interacting features found through UV observations \citep{osk01b}. \zp\ shows a 5\,d modulation also attributed to corotating regions \citep{mof81}. New, continuous X-ray observations of \zp\ would be required to better characterize the timescale of the detected variability, and to establish whether these variations can be associated with such features.

\subsection{Autocorrelation analysis}

\begin{figure*}
\includegraphics[width=7cm,bb=40 150 380 710, clip]{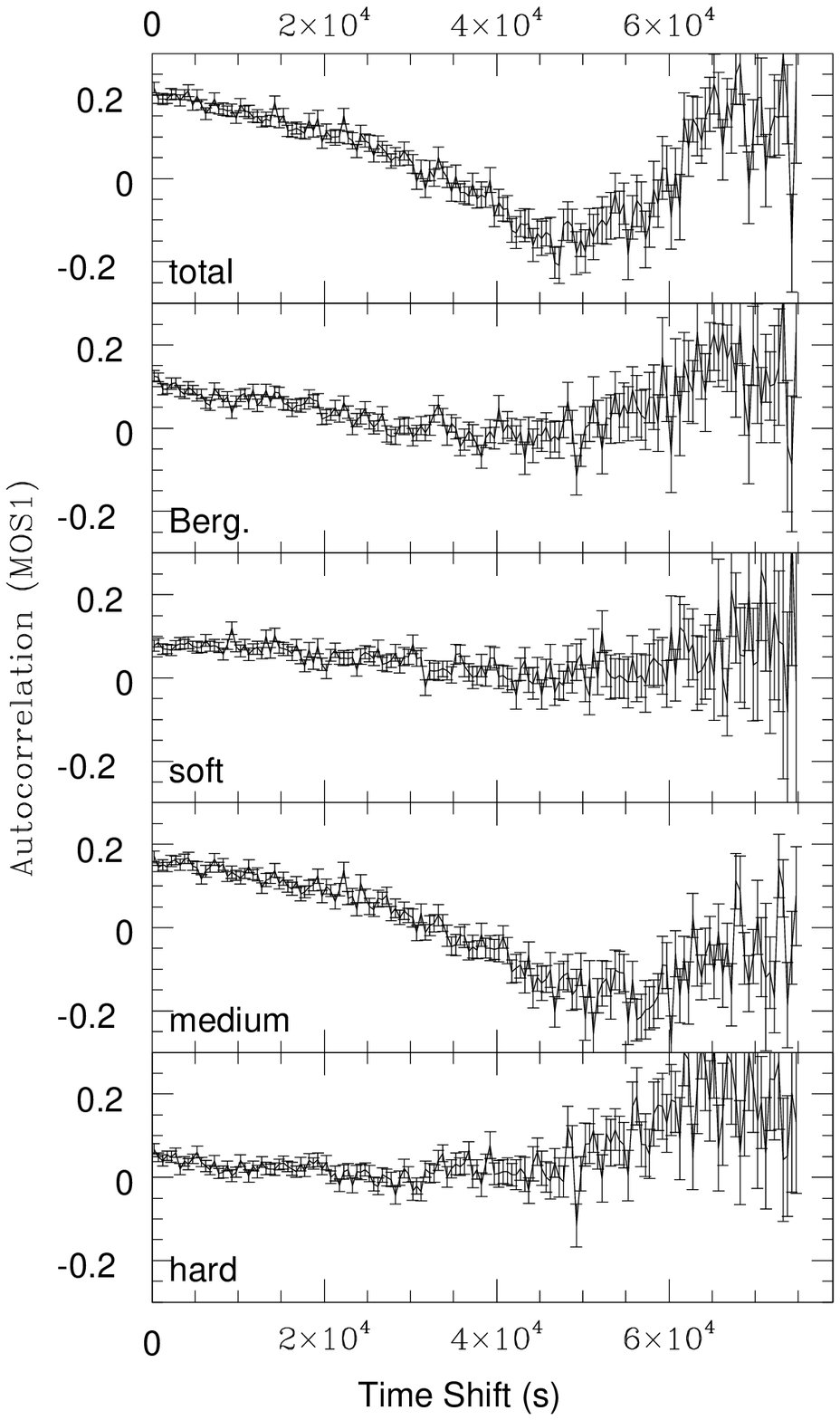}
\includegraphics[width=7cm,bb=40 150 380 710, clip]{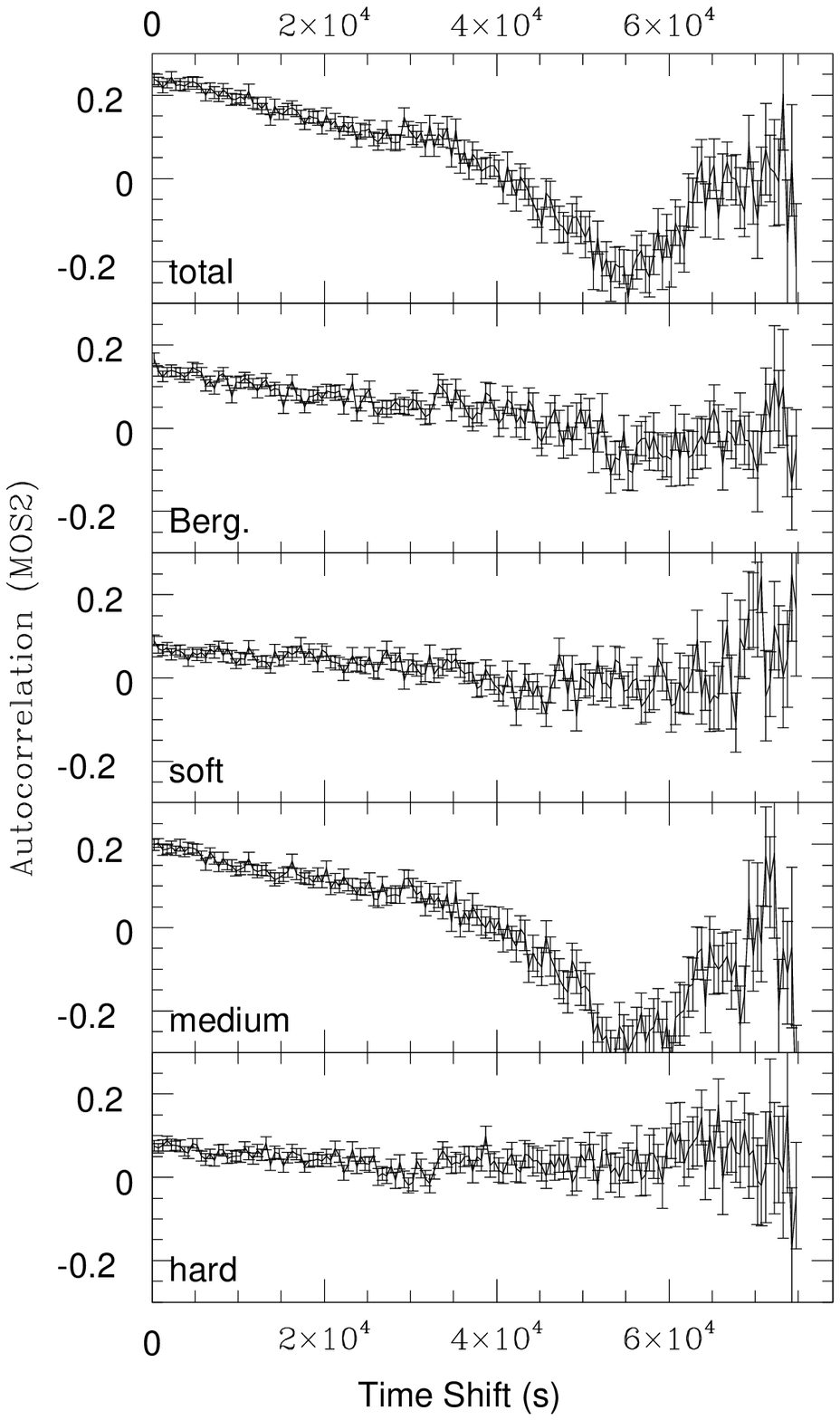}
\includegraphics[width=7cm,bb=40 150 380 710, clip]{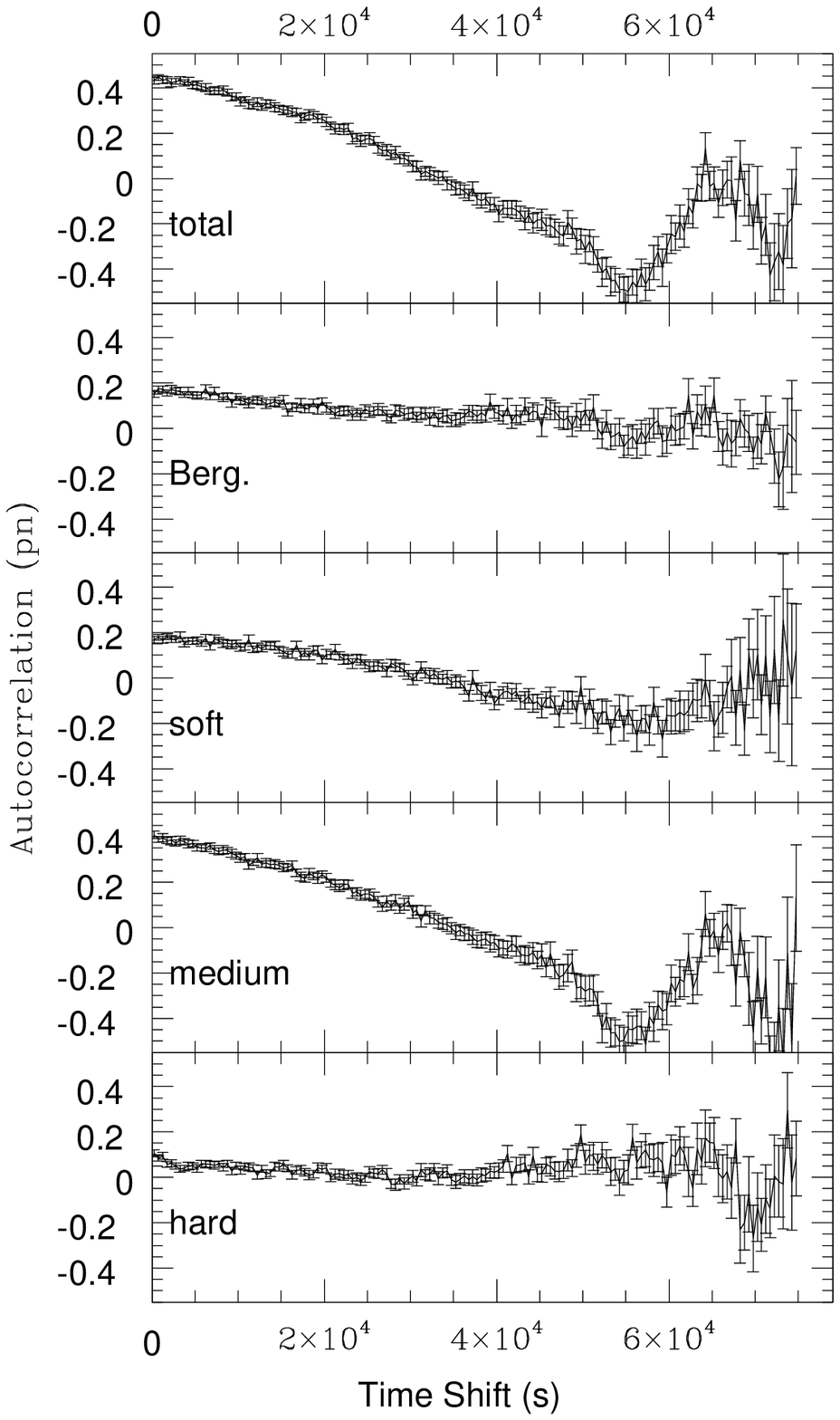}
\includegraphics[width=7cm,bb=40 255 380 710, clip]{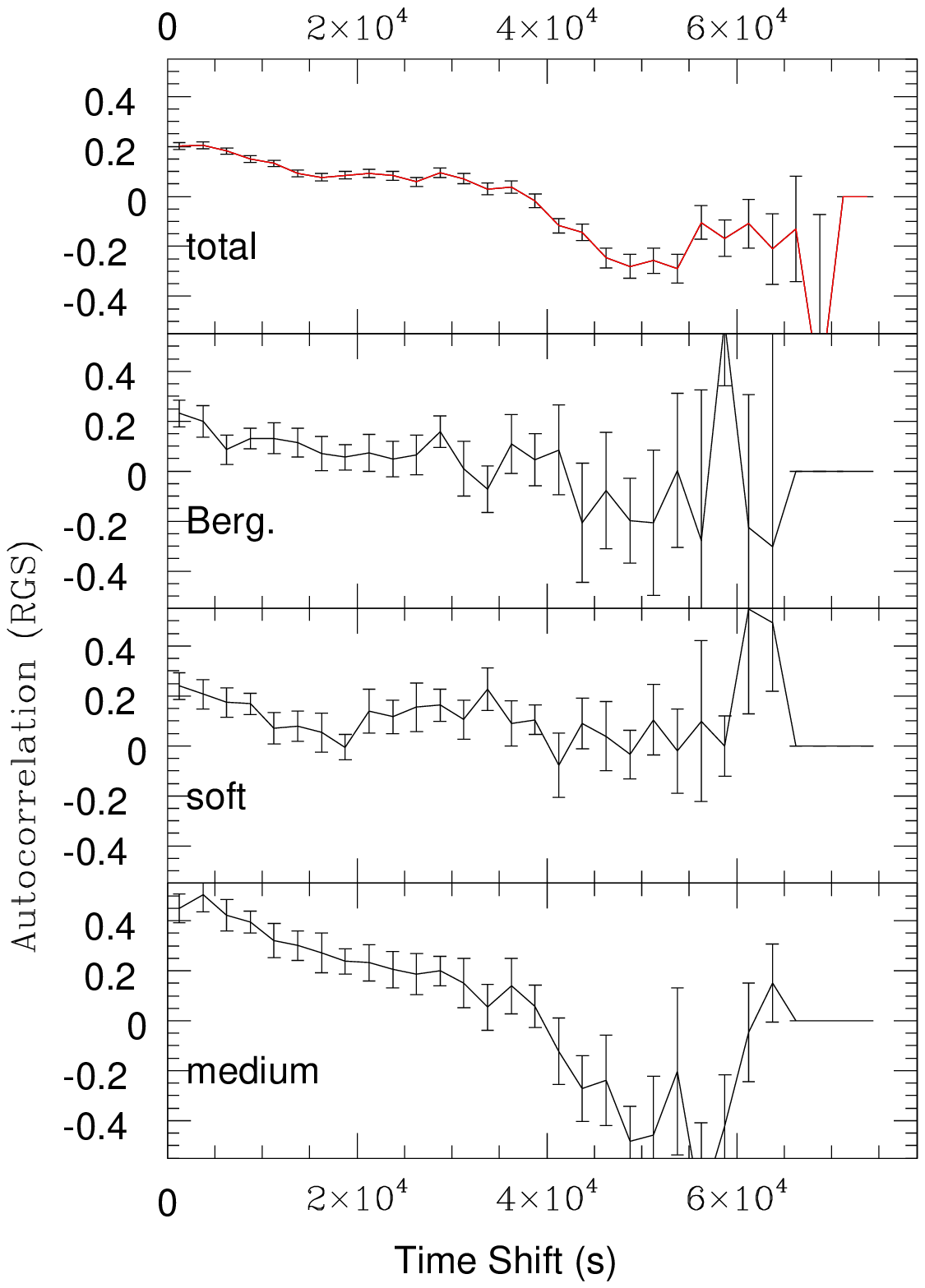}
\caption{Autocorrelation functions for the EPIC instruments and the combined RGS data.}
\label{autocorr}
\end{figure*}

To search for temporal links between the photon arrival times, the data corrected for the instrumental effect were also analyzed using autocorrelation methods \citep{ede88}. The resulting autocorrelation functions are shown in Fig. \ref{autocorr}. Note that the binned functions were normalized by the actual number of points used and, as usual, by the dispersion around the mean (observed variance $s_{ts}^2$). The former normalization is needed as the individual exposure times range from 15 to 77ks (Paper I), so that more data points can be used for some time shifts.

A slight positive correlation appears for the total and medium bands for time shifts below 20ks. A slight anticorrelation is then detected for shifts of 50-60ks, i.e. similar to the typical duration of an exposure. A similar behaviour, though with even smaller amplitude, is seen for the soft, hard and Bergh\"ofer energy bands. This slow evolution towards lower correlation values confirms the existence of weak coherent variations with timescales of tens of ks. 

The correlation amplitudes are however small, hence not highly significant, statistically speaking. However, a strong positive correlation for small shifts is expected since the wind configuration remains the same during some time (about one flow time, which is $R_*/v_{\infty}$=5.8ks for \zp, as $v_{\infty}$=2250\kms\ and $R_*$=18.6\,R$_{\odot}$, using the parameters of \citealt{osk06} and references therein).  For longer shifts, two behaviors are then possible. On the one hand, the time bins could be independent for shifts longer than a flow time, as in a fully stochastic wind. In this case, the correlation function would rapidly drop to zero. On the other hand, if the time bins are coherently linked by a shallow trend, the correlation function should slowly decrease. This decrease is indeed what is observed, but we would thus expect a correlation signal with larger amplitudes.

In practice,  however, even a perfectly correlated signal (i.e., with a correlation of 1) will be diluted by the observational Poisson noise. The expected correlation value will then be given by
\begin{equation}
C = 1 - \frac{s^2_{n}}{s^2_{ts}}
\end{equation} 
where the observed variance $s^2_{ts}$ comprises the noise and  the coherent variations of the source, whereas $s^2_{n}$ only corresponds to the noise.
For the pn lightcurve in the total band with 200s time bins (globally detrended, i.e. corrected for the instrumental effect), $s^2_{ts}$ is 0.089\,cts$^2$\,s$^{-2}$ and the variance due to Poisson noise is estimated to be around 0.049\,cts$^2$\,s$^{-2}$. The peak height $C$ should then reach 0.45, which is in good agreement with the value observed in Fig. \ref{autocorr}  for small time shifts. This implies that the intrinsic correlation of the noiseless signal could be nearing one: the trends are thus real. 

\subsection{X-ray lines}

As already mentioned in Sect. 3, RGS lightcurves were also extracted for the brightest isolated X-ray lines, enabling to determine whether these lines vary over shorter timescales. The observed variations are small, generally within 1$\sigma$ error bars. The only significant and coherent results are found for Revs. 0156 and 1814. For the former, a linear trend with a positive slope provides a much better fit to the Ne\,{\sc ix} line flux. For the latter, the Ne\,{\sc ix} line appears variable while the O\,{\sc viii} clearly increases. Note however that these changes are detected with more significance using RGS2 data than with RGS1 data. 

\begin{figure}
\includegraphics[width=9cm,bb= 50 180 565 715, clip]{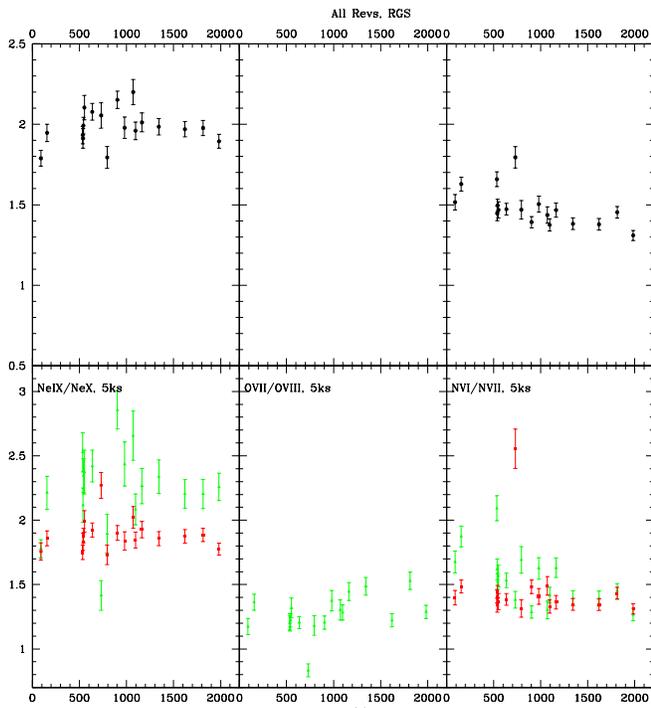}
\caption{ Ratios of count rates recorded for He-like and H-like elements (top: RGS, bottom: green triangles for RGS1, red squares for RGS2). Note that the signal for RGS is the sum of counts recorded for both RGS1 and RGS2, divided by the full exposure time. It is thus a kind of average between the two instruments (i.e. $[RGS_1+RGS_2]/2\Delta t$). However, when no signal is recorded in one instrument, like is the case for O\,{\sc vii} in RGS2,  the RGS data are not shown.}
\label{ratios}
\end{figure}

Since the individual count rates are available, we also calculated the ratios between lines of the same elements - O\,{\sc vii}/O\,{\sc viii}, Ne\,{\sc ix}/Ne\,{\sc x} and N\,{\sc vi}/N\,{\sc vii} (Fig. \ref{ratios}). These ratios reflect the temperature \citep{blu69,wal07}, and any change in the ratios would therefore be linked to temperature variations. Formally, opacity variations may also play a role in changing these ratios. However, the latter are not as sensitive to temperature as to absorption: a change in temperature by 15-20\% results in a doubling of these ratios (note that the temperature dependence is similar for all ratios), whereas change of the absorbing column of a smooth wind by a factor of 2 (from 0.1 to $0.2\times10^{22}$\,cm$^{-2}$) yields variations of 7.5\%, 25\% and 35\% in the Ne, O and N ratios, respectively. As for count rates, the ratios were tested using $\chi^2$ tests. For the individual exposures, only two features appear significant. For Rev. 0156, the Ne\,{\sc ix}/Ne\,{\sc x} is better fit by a line with negative slope. Similar but shallower possible trends are detected by eye for the two other ratios, but they are not formally significant. For Rev.\ 1814, the Ne\,{\sc ix}/Ne\,{\sc x} appears much better fit by a concave/U-shaped parabola, but no similar signal is seen for the two other ratios. Note that these two revolutions are amongst the variable ones (see above), and that the reported changes may be linked to slight changes in spectra (see Sect. 4.2), for which higher-sensitivity instruments are needed for a detailed characterization.

\section{Short-term variations (hours)}

Once the long-term instrumental decrease and the daily trends are removed, what signal is left? In principle, this is where the variability due to embedded wind-shocks should appear. In what follows, we will first present the new observational results, then the theoretical predictions of \zp\ variability, and finally compare the two to derive constraints on the wind structure.

\subsection{Observed lightcurves}

Relative dispersions were calculated for each of the observed lightcurves. These lightcurves have two specific features which do not exist in the models (see next subsection): a temporal binning and a binning over a range of energies. In principle, this somewhat smoothes out any variability if it is present with timescales shorter than the time-bin length, or if it changes strongly with energy (e.g. the lightcurves at 0.3 and 0.4\,keV being uncorrelated). However, the Poisson noise inevitably  impacts on the data and therefore prevents us to detect low-level variability (i.e. a few percent) with time bins smaller than 200s (see last line of Table \ref{dispdata}) - this limit would only be changed if a more sensitive instrument was used. In addition, the emitting parcels contain plasma emitting over a range of energies, even if isothermal, and this implies some correlation over different ranges of energies. A comparison with synthetic curves can thus provide useful insights into the structure of the wind. 

As mentioned below, the steps in the synthetic lightcurves correspond to different wind configurations. The dispersion calculations for the data were thus performed in the following way: (1) the original lightcurves of each exposure were first detrended using the best-fit linear trend derived from $\chi^2$ calculations (see above) as the simple model (presented in next section) does not predict  nor model such features, (2) the count rate values of the 200s lightcurves were then sampled at 5 and 10\,ks intervals (corresponding to about one and two wind flow times, respectively), i.e. only considering every 25th or 50th value, and (3) the relative dispersions of these new, reconstructed lightcurves were finally evaluated using
$$rel.\, disp=\left[ \sqrt{ \sum(CR_i-a-b\times t_i)^2/(N-2)}\, \right] \, /mean$$
where $a$ and $b$ are found from the best-fit trend, see Sect. 3.2.1, and $mean=\sum (CR_i/\sigma_i^2)/\sum(1/\sigma_i^2)$. This enables us to try to reproduce the `independent wind configurations' of the successive `time steps' in the synthetic curves. Dispersions were also calculated for the full  (i.e. considering all time bins) 200s, 500s, 1ks and 5ks lightcurves (after detrending). When one compares the dispersions of the full 200s lightcurves to these of the reconstructed lightcurves, they appear very similar.  The maximum differences amounts to $\pm$4\% in the worst cases, or smaller ($<$1\% in absolute value) when the observations are longer. This is unsurprising since dispersion estimates on, e.g., 2 or 3 bins being less precise than on e.g. 50 bins: when one samples the lightcurves every 25 or 50 steps, a particularly discrepant realization of the noise has more impact when there are less bins. The sparse sampling therefore does not change much the results, so Table \ref{dispdata} reports only the dispersions of the full lightcurves. This Table gives in first column the revolution number, in columns 2--6 the dispersions measured for the main 5 energy bands and the 200s binning, and in the 7th column the number of data points used; Columns 8--13 report similar values, but for the 5ks binning. The last line provides for comparison the expected Poisson error, relative to the mean (i.e. $1/ \sqrt{ct\,rate \times bin\,length}$). Note that only the EPIC data were used here, as the noise of the RGS lightcurves is even larger and would more easily mask subtle variations. 

Looking at Table \ref{dispdata}, one thing is obvious: despite the fact that the exposure-long trends mentioned above are often not perfectly linear, the measured relative dispersions around the best-fit lines are close to that expected on the sole basis of the Poisson noise. This means that the true short-term variability of \zp\ - and its potential wavelength dependence - remains hidden in the noise, implying a very small amplitude for these `intrinsic' short-term changes. 

These results should now be translated into constraints on the wind structure.  Considering X-rays to be emitted by many hot parcels, one may naively think that it is sufficient to notice that a 1\% dispersion ``naturally'' corresponds to 10$^4$ ``emitters'', using simple Poisson statistics. However, reality is not as simple: emitters suffer from different amounts of absorption depending on their location and, worse, the absorption may be clumped too. In both cases, the simple reasoning completely fails to apply, and a dedicated model is thus needed.

\subsection{A simple wind model}

Embedded wind-shocks are considered as responsible for the X-ray emission of O-type stars. In these expanding and unstable winds, fast material encounters slow-moving material, giving rise to zones of dense gas, and the mutual collisions of such dense features then give rise to the X-rays. The radiation hydrodynamic models (e.g. \citealt{fel97a,fel97b}) predict that the collision, and thus the heating, occurs quickly (tens of seconds), and the cooling of the resulting hot plasma is also rather rapid, as the cooling time remains lower than the wind flow time for distances of several tens of stellar radii. A strong X-ray variability (2 orders of magnitude in amplitude) was predicted. However, early $ROSAT$ and {\it Einstein} observations failed to show such a strong variability of X-ray fluxes from massive stars \citep[and references therein]{ber94b}. To reconcile the results of the early observations with wind-shock theory, it was suggested that lateral break-up of the shells can lead to the presence of many parcels, resulting in low variability \citep{cas83,fel97a}: therefore 2-D or 3-D models were needed.  First attempts for 2D wind models were made by \citet{des03,des05} but only isothermal winds (i.e. without X-ray generation) have so far been considered. Another example is the recent work of \citet{cas08} on bowshocks, but the model is not yet self-consistent in terms of production and evolution for an ensemble of clumps. Unfortunately, no new, multi-D hydrodynamical model has thus tackled the problem of X-ray generation in stellar winds, and we are thus left with alternative modelling paths.

In this context, \citet{osk01a} considered a spherically symmetric smooth cool wind permeated with discrete zones of hot X-ray emitting gas. It was found that X-ray variability depends on the frequency with which hot zones are generated, and on the cool wind opacity for the X-rays. It was shown, that in such smooth cool wind, the variability in soft band is expected to be smaller than in the hard band. \citet{osk04} further developed a 2-D wind model where not only hot parcels are discrete, but the cool absorbing wind can be clumped too. We apply this model in the present paper to investigate the resulting X-ray variability. We only briefly recall here its basic features. 

Our model was designed to correctly perform the radiative transfer of X-rays in stellar winds, and it also reproduces the features derived from RHD simulations. In this model, the X-ray emission originates from discrete zones of hot gas randomly distributed in an X-ray production zone extending from 1.2 to 100 $R_*$. The choice of such a large zone may at first seem surprising, since the prediction of the 1D hydrodynamical modelling place the X-ray emitting regions within a few stellar radii of the stellar surface. However, a detailed {\it global} analysis of the high-resolution spectrum of \zp\ (Herv\'e et al., submitted) shows that the X-ray emission zone must actually extend up to $\sim85 R_*$ to reproduce the observed spectrum, hence our choice of a large emission zone.  Note that the lower boundary of the region was chosen taking into account the analyses of individual fir triplets, which place onset radii in that range \citep[e.g.][ - note that these analyses also consider large emitting zones, with emission formally extending to infinity]{wal07}. All emitting parcels of gas contain the same amount of matter. The line emission is powered by collisional excitation and therefore scales with the density-squared. The density of the wind is derived based on the stellar mass-loss rate: from the mass conservation, it is $\rho(r) = \dot M / [4 \pi r^2 v(r)]$ where $\dot M$ is the mass-loss rate and the wind velocity $v(r)$ is $v_{\infty} (1-R_0/r)^{\beta}$ [with $\beta$=1 and $R_0$ chosen so that the photospheric velocity $v(R_*)=0.01 v_{\infty}$], as commonly used for a massive star wind.  During motion, each hot fragment expands according to the continuity equation. Hence, the intrinsic unattenuated X-ray luminosity of each hot fragment scales as $1/r^2 v(r)$. The probability to find a hot fragment in the radius interval [r, r+dr] scales with $1/v(r)$, i.e. emitters are concentrated at inner radii, where the wind is slow. The random radial location of fragments is determined by von Neumann's rejection method \citep[e.g.][]{pre92} and their angular distribution is also random, with a uniform distribution over the sphere. Absorption and emission are decoupled: there is no self-absorption for the emitting material and no re-emission of X-rays after absorption. The model allows for further sophistications, i.e. emissivity can have a different scaling with density and density can also be a parameter. However, these parameters affect the shape of X-ray emission line profiles, not relative variability, so in present study, we use the simplest emissivity. 

\begin{figure}
\includegraphics[width=8cm]{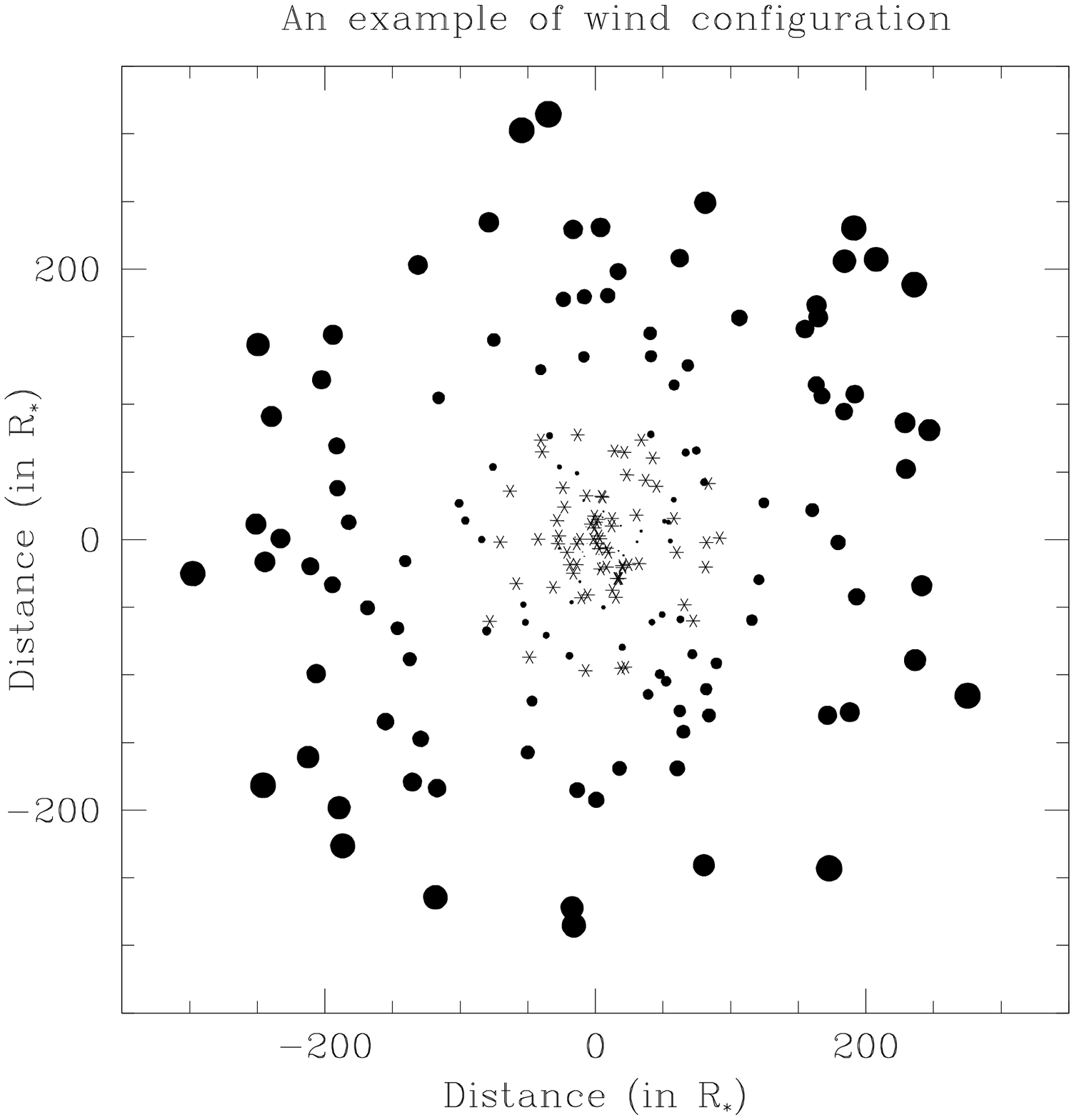}
\includegraphics[width=8cm]{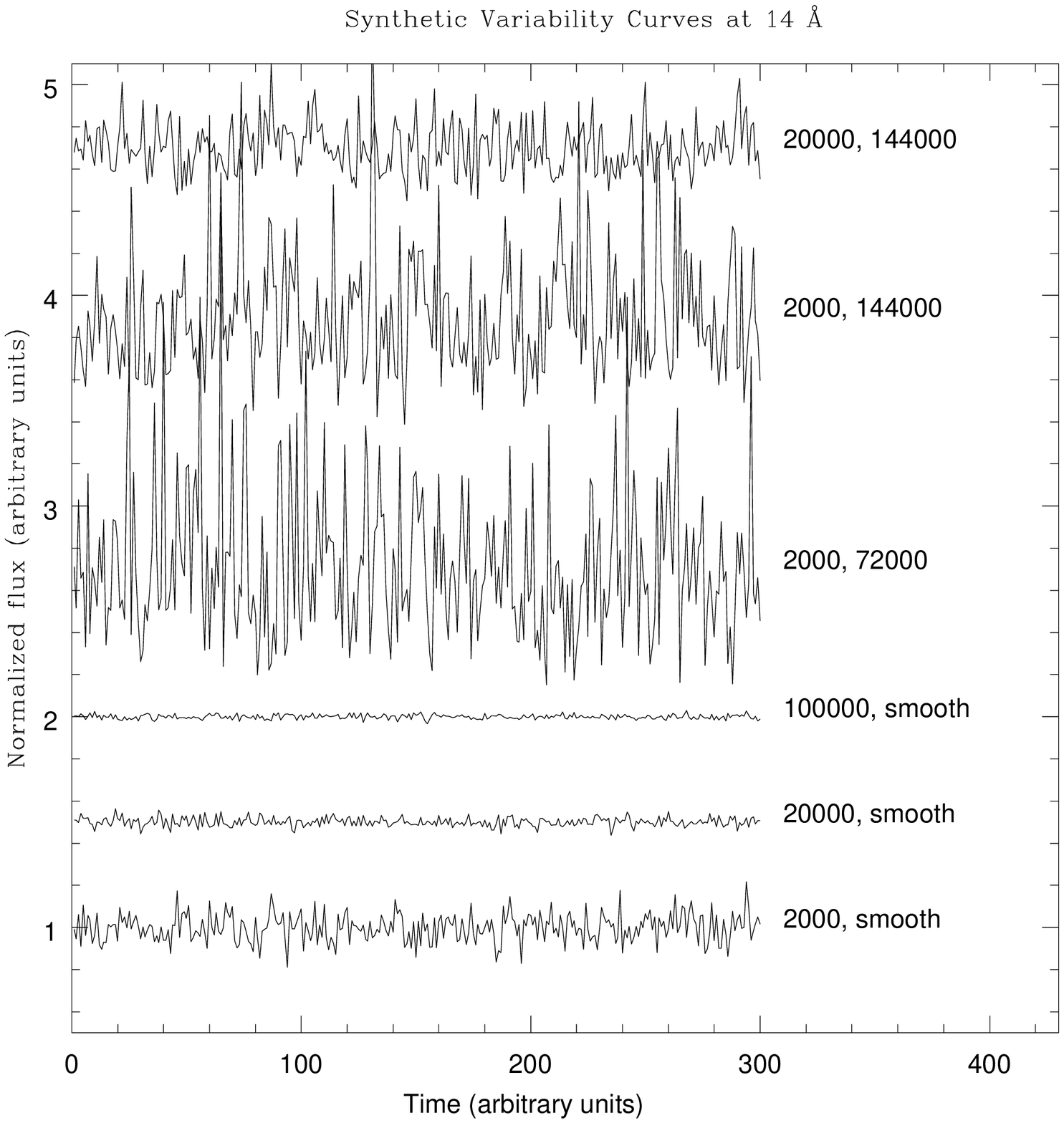}
\caption{{\it Top:} A snapshot showing the wind structure  in the model for a small number of parcels. Black dots represent the absorbing clumps (which, despite their color, are not totally optically-thick), while crosses correspond to the emitting regions. Note that the size increases with distance to the star, as the parcels keep their angular size, as seen from \zp. Only a small number of clumps is shown, for clarity. {\it Bottom:} Excerpts of synthetic variability curves at 14\AA\ for a set of wind configurations. Next to each synthetic lightcurve are indicated the numbers of emitting zones followed by the number of absorbing clumps. For clarity, the lightcurves have been vertically shifted.}
\label{model}
\end{figure}

Once produced, the X-ray emission propagates through the cooler stellar wind which can absorb it. The velocity of this cool wind is assumed to follow the same velocity relation as for the hot wind component (the so-called "beta-law", see above). The cool stellar wind is assumed to be either a smooth cool wind or a set of cool clumps (randomly distributed over the 1.5 to 316 $R_*$ range), both cases having the same overall mass-loss rate and optical-depth. For the fragmented wind model, we use the "cones" model of \citet{osk04}, with a lateral extent of 1 degree for the spherical absorbers. A sketch of the model geometry is shown in Fig. \ref{model}. Random radii are generated for each cone using the $1/v(r)$ probability and von Neumann's rejection method, again. The total mass of a homogeneous wind enclosed between two subsequent radii is considered to be swept up in a dense fragment with the same optical depth as the homogeneous wind material, so that the fragment location is given by 
$r^2=[\int_a^b dr'/v(r')]/[\int_a^b dr'/{r'^2 v(r')}]$ where $a$ and $b$ are two subsequent random radii in the set determined by von Neuman's method. Note that, depending on the set of radii, cool clumps do not have all the same density, some being optically-thick while others are not blocking much light.

The wind parameters ($v_{\infty}$, $\dot M$, abundances) are derived from a non-LTE atmosphere model specific to \zp\ \citep{osk06}, matching the optical/UV spectrum of the star. All  these fix the clump location, size, density and optical depth, leaving as only free parameter the number of hot and cool clumps. It may be noted that this model reproduces well the observed X-ray line profiles, despite its simplifications \citep{osk04}.

Our analysis of observational data did not reveal strong spectral trends in the variability. Therefore, for simplicity and clarity we simulate here only the monochromatic X-ray flux at a few selected representative wavelengths: 6\AA\ ($\sim$2\,keV, midpoint of the total EPIC band), 14\AA\ ($\sim$0.9\,keV, midpoint of the medium EPIC band), and 19\AA\ ($\sim$0.65\,keV, midpoint of the total RGS band). 

Some comments should be made about the synthetic lightcurves. First, the calculated flux is monochromatic and in arbitrary units. Only relative dispersions, for example, can be calculated and compared to the data. Second, these variability lightcurves are not, strictly speaking, function of time. The time-dependent radiative hydrodynamic simulations of \citet{fel97b} show that on the timescale longer than the flow time (5.8\,ks for \zp, as $v_{\infty}$=2250\kms and $R_*$=18.6\,$R_{\odot}$, see \citealt{osk06} and references therein), the wind structure is renewed and is independent of the previous wind configuration. Our model reproduces this situation: each realization of our 2-D stochastic wind model is independent from the previous one and each model run represents the wind configuration on a timescale shorter than the cooling time. In other words, we model the wind at some arbitrary moment of time, and compute the X-ray emergent flux at this moment; the next data point is calculated for a randomly different wind configuration for both absorbing and emitting parcels. We do not follow the wind expansion (this will be the subject of a paper by Oskinova et al., in prep). 
While this approach does not allow us to model the detailed time evolution of the X-ray flux on timescales of hours, it allows to model the relative amplitude of the X-ray variability for long stretches of randomly distributed observations, as appropriate for this \xmm\ observing campaign. Examples of such `lightcurves' are shown in the right panel of Fig. \ref{model},  while relative dispersions are presented in Table \ref{dispmodel}.

Several conclusions can be drawn from Table \ref{dispmodel}:
\begin{itemize}
\item As the number of emitting or absorbing parcels increases, the variability decreases, as could be expected. However, it is important to note that dispersions of one percent are only found when these numbers approach $10^5$. 
\item A smooth wind is less variable than an otherwise equivalent clumped wind.
\item As wavelength increases (or energy decreases), the relative dispersion for a clumped wind remains stable or slightly increases whereas it decreases for a smooth wind (as already reported in \citealt{osk01a}). In a clumped wind, the variability is thus less energy-dependent than in a smooth wind - in the extreme limiting case of a wind consisting of only opaque X-ray clumps, no energy dependence is expected \citep{osk04,osk06}. 
\item If absorbing clumps are distributed over a smaller region ($R_{\rm max}$ of 100 rather than 316$R_*$), then the variability decreases. This is the effect of a smaller radial separation between clumps.
\end{itemize}

Comparing these theoretical predictions with the observational results (see previous subsection), we found that the number of emitting and absorbing parcels is huge. Indeed, even in the most favourable case of smooth cool wind absorption, which is the least variable case, more than 100\,000 hot X-ray emitting zones must be present  and contribute to the X-ray emission so that the relative flux variations remain below 1\%. This number further increases when the cool wind fragmentation is also included in the model. 

It must be underlined that no previous study has put direct constraints on the number of clumps in O-stars. Some studies exist, however, for evolved massive stars. For example, \citet{lep99} report that ``between $10^3$ and $10^4$ clumps in the line emission region are needed to account for the line profile variability of the WR stars'' that they analyzed. The X-ray data of \zp\ suggest an even larger number, though this should apply to a larger zone than formation regions of optical lines. In addition, \citet{dav07} explained the polarization level of LBVs by either a few massive, optically-thick clumps (ejection rate of $\lesssim0.1$\,clump per flow time) or many small, optically-thin clumps (ejection rate of $\gtrsim10^3$\,clumps per flow time), with the latter option usually favored in literature.  Considering that the X-ray emission region and the cool wind absorption region cover several tens of stellar radii, our conclusion appears consistent with Davies' result, despite the different nature of the objects under consideration.

Thus, the sensitive XMM observations reveal that the stellar winds of O-stars are highly structured on small scales. It remains to be seen whether the theory of stellar wind instability can explain such a high degree of fragmentation. Some 2D models of the line driven instability (LDI) in isothermal\footnote{The wind temperature was set at the effective temperature of the star.} stellar winds ``show that radially compressed shells that develop initially from the LDI are systematically broken up by Rayleigh-Taylor or thin-shell instabilities as these structures are accelerated outward'' \citep{des03}, hence producing a lot of small-scale 2D structures, but the same authors later found ``lateral coherence of wind structures'' \citep{des05} and the question of the size and number of clumps therefore remains unsettled. The hydrodynamical simulations of \citet[to this day the sole ones that tackled the problem of X-ray generation, though in 1D]{fel97a} predicted that only a few strong shocks, simultaneously present in massive star winds, are responsible for  most of the X-ray emission, whereas our results strongly challenge this. The lateral break-up of the X-ray emitting shells, advocated by \citet{fel97a} for lowering the X-ray variability, cannot provide an explanation (see Table \ref{dispmodel}). Moreover, if clumping is induced by sub-surface convection, hydrodynamical stellar evolution codes indicate that the total number of clumps in O-stars should typically amount to $6\times10 ^3-6\times 10^4$ \citep{can09}. Our data indicate  an order of magnitude larger values, therefore prompting further investigation on the nature of radiatively-driven stellar winds. 

\section{Summary and Conclusion}

We have analyzed an exceptional set of X-ray observations of \zp : to date, there exists no more sensitive dataset nor a dataset with a better (time and spectral) coverage for a massive star. 

Over the decade of observations of \zp, a decreasing trend is clearly seen in the count rates. As a comparison with the fluxes determined from spectral fits (see Paper I) and with other X-ray sources shows, this is mostly due to instrumental/calibration problems (probably the ageing detector, whose sensitivity decreases with time), and is not yet taken into account in the data calibration process. 

Comparing the X-ray lines appearing in the 18 available high-resolution spectra yields again no significant, true long-term changes. Some shallow, 1$\sigma$ line profile variations are however reminiscent of those seen in optical - more sensitive observations should confirm this, and  pinpoint the timescales on which that occurs. 

Once the instrumental effect is taken out, we do not detect flare-like bursts of X-ray emission nor short-term variations ($<$1d, like e.g. the stellar pulsations detected in the optical domain) in the individual exposures. However, we detect statistically-significant variations of the count rates of \zp\ on timescales of $>1$\,d. Indeed, the lightcurves with the longest time bins appear more variable than those with shorter bins, and the longest datasets are systematically found to be variable. The detected changes appear as shallow increasing (Revs. 0156, 1071, 1620, 1814, and 1983) or decreasing (Rev. 1343) trends, or a mix of both (Rev. 1164).  No clear dependence with energy is found: in particular, the hard band is not the most often variable one. This suggests that the bulk of the X-ray emission is affected, not only a high-energy emission tail linked to phenomena such as magnetic confinement. Furthermore, no evidence for a coherent, systematic periodicity is found either. These slow modulations cannot be explained by the embedded wind shock scenario but are consistent with the presence of large-scale, slowly-moving structures in the wind, which may, for example, result from corotating interaction regions in the wind. Clearly, more data, specifically covering the full rotation period, are needed to settle the question of the origin of such features.

Once instrumental and daily trends are taken out, \zp\ shows a surprisingly low level of variability. This places stringent constraints on the wind structure. A wind variability  model tailored to \zp\ has been undertaken. It shows that only a very large number of emitting and absorbing clumps ($>10^5$)  are able to reproduce the observed lightcurves of \zp. This is the first time that such a limit  has been placed for an O-type star. The stringent limit on number of clumps that we established questions some results from existing models of stellar winds (link clumping-convection, fragmentation level) and X-ray production (number of hot gas zones), but these models were not calculated in 3D. Future model developments should be done, taking into account our high clumpiness result.

\begin{acknowledgements}
YN and EG acknowledge support from the Fonds National de la Recherche Scientifique (Belgium), the Communaut\'e Fran\c caise de Belgique, the PRODEX XMM and Integral contracts. YN also acknowledges the \xmm\ helpdesk for interesting discussions about the data. LMO acknowledge support  by DLR grant 50\,OR\,1101 (LMO). ADS and CDS were used for preparing this document. The authors acknowledge the editor and referee.
\end{acknowledgements}

\clearpage

\clearpage

\clearpage

\clearpage

\clearpage

\clearpage

\clearpage

\clearpage

\clearpage
\begin{table}
\caption{Relative dispersions (in \%) measured  for the observed \xmm\ lightcurves after detrending (see text for detail). For the pn, the first five lines correspond to the data taken with the Medium filter (i.e. influenced by pile-up).}
\label{dispdata}
\centering
\begin{tabular}{l|cccccc|cccccc}
            \hline\hline
Rev & \multicolumn{6}{|c|}{200s binning} & \multicolumn{6}{|c}{5ks binning}\\
 & total & soft & medium & hard & Berg. & N & total & soft & medium & hard & Berg. & N \\
\hline
MOS1 \\
0156 & 5 & 10 & 7 & 12 & 8 &   189 & 1.4 & 3   & 1.6 & 2   & 1.3 &     8\\
0535 & 5 & 11 & 7 & 12 & 8 &   211 & 0.7 & 1.5 & 0.9 & 1.0 & 1.3 &     8\\
0538 & 6 & 13 & 7 & 12 & 9 &   163 & 0.8 & 3   & 1.0 & 3   & 1.2 &     7\\
0542 & 5 & 13 & 7 & 13 & 9 &   217 & 1.3 & 2   & 0.7 & 5   & 2   &     9\\
0636 & 5 & 13 & 7 & 13 & 10&    95 & 1.2 & 1.2 & 1.5 & 2   & 2   &     4\\
0795 & 6 & 13 & 8 & 14 & 9 &    96 & 0.1 & 2   & 1.7 & 4   & 1.8 &     4\\
0903 & 6 & 13 & 7 & 15 & 9 &   109 & 0.7 & 4   & 2   & 1.8 & 1.2 &     4\\
0980a& 6 & 14 & 7 & 11 & 9 &   147 & 0.4 & 2   & 1.1 & 1.3 & 1.1 &     6\\
0980b& 6 & 13 & 7 & 12 & 8 &    69 & 0.4 & 1.9 & 1.3 & 3   & 0.3 &     3\\
1096 & 5 & 13 & 7 & 13 & 9 &   238 & 1.2 & 3   & 1.9 & 3   & 2   &    10\\
1164 & 6 & 12 & 8 & 12 & 9 &   203 & 1.9 & 2   & 2   & 4   & 2   &     8\\
1343 & 6 & 13 & 8 & 13 & 9 &   243 & 1.9 & 3   & 2   & 2   & 3   &    10\\
1620 & 6 & 13 & 8 & 13 & 9 &   275 & 1.2 & 3   & 1.6 & 1.9 & 1.8 &    11\\
1814 & 6 & 14 & 7 & 13 & 9 &   320 & 1.8 & 4   & 1.7 & 3   & 2   &    13\\
1983 & 6 & 13 & 8 & 13 & 9 &   383 & 1.5 & 3   & 1.7 & 4   & 3   &    15\\
\hline
Poisson   & 5 & 12 & 7 & 11 & 8 & & 1.0 & 2 & 1.4 & 2 & 1.6 & \\
\hline
\hline
pn \\
0156 & 3 & 5 & 4 &  8 & 5 &   175 & 0.8 & 0.9 & 1.0 & 1.0 & 1.3 &     7\\
0535 & 3 & 4 & 4 &  9 & 6 &   122 & 0.2 & 0.4 & 0.4 & 3   & 0.6 &     5\\
0538 & 2 & 5 & 3 &  7 & 4 &    65 & 1.2 & 2   & 0.7 & 0.9 & 1.7 &     3\\
0542 & 2 & 5 & 3 &  7 & 5 &   111 & 0.3 & 1.2 & 0.7 & 1.0 & 0.5 &     4\\
0552 & 3 & 5 & 4 &  7 & 4 &    61 &     &     &     &     &     &     2\\
0535 & 3 & 5 & 3 &  9 & 5 &    65 & 1.1 & 2   & 0.9 & 1.5 & 0.9 &     3\\
0538 & 3 & 5 & 4 &  9 & 6 &    79 & 0.5 & 0.4 & 0.4 & 1.9 & 0.8 &     3\\
0542 & 3 & 6 & 5 &  8 & 6 &    70 & 0.6 & 0.2 & 0.2 & 3   & 1.9 &     3\\
0636 & 4 & 6 & 4 & 10 & 7 &   168 & 0.6 & 1.7 & 0.9 & 1.4 & 1.2 &     6\\
0731 & 3 & 6 & 4 & 10 & 6 &   173 & 0.5 & 0.6 & 0.7 & 1.6 & 1.3 &     7\\
0795 & 3 & 6 & 4 &  9 & 6 &   124 & 0.9 & 0.8 & 1.1 & 0.6 & 1.4 &     5\\
0903 & 3 & 7 & 5 &  9 & 7 &   149 & 1.2 & 0.7 & 1.8 & 3   & 2   &     6\\
0980 & 4 & 7 & 5 & 10 & 7 &   158 & 0.8 & 0.8 & 0.6 & 1.4 & 1.9 &     6\\
1071 & 4 & 6 & 5 & 11 & 6 &   111 & 0.6 & 0.4 & 1.0 & 1.3 & 1.0 &     4\\
1096 & 3 & 6 & 4 & 10 & 7 &   235 & 0.5 & 0.7 & 0.9 & 3   & 1.6 &     9\\
1343 & 4 & 6 & 5 & 10 & 7 &   244 & 0.6 & 1.1 & 1.0 & 1.8 & 1.0 &     9\\
1620 & 4 & 6 & 5 &  9 & 6 &   273 & 1.0 & 1.2 & 1.3 & 1.7 & 1.5 &    11\\
1814 & 4 & 7 & 5 & 10 & 7 &   319 & 1.8 & 2   & 2   & 2   & 1.9 &    13\\
1983 & 4 & 7 & 5 & 10 & 7 &   382 & 1.5 & 1.3 & 1.9 & 4   & 3   &    15\\
\hline
Poisson   & 3 & 5 & 4 & 8 & 5 & & 0.6 & 1.0 & 0.7 & 1.5 & 1.0 & \\
\hline
\end{tabular}
\end{table}

\clearpage
\begin{table}
\caption{ Relative dispersions measured  for a set of synthetic lightcurves.}
\label{dispmodel}
\centering
\begin{tabular}{lccc}
            \hline\hline
$\lambda$ & \# of abs. clumps & \# of emit. clumps & Rel. dispersion (\%)\\
\hline
6\AA  & smooth & 2000  & 16\\
6\AA  & 144000 & 2000  & 23\\
14\AA & smooth & 2000  & 6.9\\
14\AA & smooth & 5000  & 4.2\\
14\AA & smooth & 20\,000 & 2.1\\
14\AA & smooth & 50\,000 & 1.4\\
14\AA & smooth & 100\,000& 1.0\\
14\AA & 72000  & 2000  & 33\\
14\AA & 144000 & 2000  & 27\\
14\AA$^a$ & 144000 & 2000  & 16\\
14\AA & 144000 & 20\,000 & 12\\
19\AA & 144000 & 2000  & 25\\
19\AA$^b$ & 288000 & 2000  & 26\\
19\AA & 144000 & 20\,000 & 12\\
\hline
\end{tabular}
$^a$ In all other runs the smooth wind starts after $316\,R_*$, while in this run the cool fragment zone of the wind ends at $100\,R_*$, which decreases the variability. \\
$^b$ In this run, the number of clumps in the radial direction is the same (400) as in the previous case but the lateral size of each clump is 0.5$^{\circ}$. No strong impact on the variability is found.\\
\end{table}

\clearpage

\begin{appendix}
\section{Results for individual exposures}
In this appendix, we present the variability properties of each exposure. We also remind the reader about background flares affecting the exposure (though they have been cut out during the processing, see Paper I).  
\subsection{Rev. 0091}
This dataset only comprises data from the RGS. There were narrow background flares scattered all over the exposure and a broad flare affected the mid-exposure. At a significance level of 1\%, no trend or deviation from a constant is detected. Count rates may be higher towards the end of the exposure, but only at the 10--20\% level. A feeble oscillation can be detected by eye in the Bergh\"ofer band, with a recurrence time of about 10ks, but it does not reach an amplitude larger than that of the 1-$\sigma$ error bars. The count rate in the medium band and with the longest time bin appears as the most variable (i.e. they show a $\chi^2$ associated with the lowest significance level, but this significance level is not $<$1\%).

\subsection{Rev. 0156}
A narrow background flare affected the pn data at mid-exposure. This dataset shows a trend towards increasing count rates. Though less obvious in MOS2 and RGS1, this trend is clearly detected through the significant improvement of the $\chi^2$ when using a linear rather than a constant fit. This trend is detected for the total, soft, and medium bands (EPIC), and for the medium band (RGS), i.e. it concerns photons with energies below 1 keV.  For the EPIC-MOS1 lightcurve in the total band, the increase rate is 1.7$\pm$0.7$\times 10^{-6}$ cts\,s$^{-2}$ (i.e. a 3\% increase over the $\sim$40ks exposure). For EPIC, count rates in the total band and/or calculated using the longest time bin appear as the most variable; for RGS, the medium band data and/or the lightcurves with the longest time bins are the most variable.

\subsection{Rev. 0535}
No soft proton flare is detected for this exposure. Data are compatible with a constant count rate, without any trend detected or any obvious oscillation. The pn observation is cut in two parts for this revolution and the next two (see Paper I): here, the second pn dataset (taken with Medium filter) appears more variable than the first part (taken with Thick filter), in all bands and for the smallest time bin, but this difference is not formally significant. For EPIC, the hard band lightcurves (esp. those obtained with the smallest time bin) are the most variable; on the contrary, for the other bands, the lightcurves calculated using the longest time bins are the most variable. RGS data do not yield any coherent result as to which time bin and energy band is the most variable.

\subsection{Rev. 0538}
A large flare occurred near the end of the observation. Data are compatible with a constant, without any trend detected or any obvious oscillation. For EPIC, the hard band lightcurves (esp. those obtained with the longest time bin) as well as lightcurves taken with the extreme time bins (200s and 5ks) are the most variable; for RGS, the shortest time bins generally yield the most variable lightcurves.

\subsection{Rev. 0542}
A large flare occurred near the end of the observation, it did not affect the MOS data. Data are compatible with a constant count rate, without any trend detected or any obvious oscillation. The first pn dataset appears more variable than the second part, in all bands for the 1ks bin, but it is not highly significant. Bergh\"ofer's band appears as the most variable for both EPIC and RGS; the longest time bins yield the most variable RGS lightcurves.

\subsection{Rev. 0552}
Only pn data are available for EPIC. A large flare occurred during the second half of the observation. The data from pn and RGS1 indicate a slight increase (linear trend better than a constant at the $<$5\% level) of the count rates in the second half of the observation, but this is not corroborated by the RGS2. While there is no time bin or energy band favoring the variability properties of the RGS data, the medium band data and lightcurves calculated using the longest time bins (except for the hard band) yield the most variable lightcurves in the EPIC-pn.

\subsection{Rev. 0636}
A large flare occurred at the end of the MOS datasets, or in the middle of pn and RGS datasets. Data are compatible with a constant, without any trend detected or any obvious oscillation. There is no time bin or energy band favoring variability.

\subsection{Rev. 0731}
A large flare occurred during the last third of the observation. Only pn data are available for EPIC. Data are compatible with a constant count rate, without any trend detected or any obvious oscillation. For RGS, the medium band lightcurve calculated with the 2ks bin appears as the most variable; for EPIC, the hard band data are the most variable except for the longest time bins.

\subsection{Rev. 0795}
A large flare occurred during the second half of the observation. The count rate recorded in the hard band is not compatible with a constant - it shows a shallow decreasing trend - but only at significance levels of 1--10\% (i.e. not formally significant). For EPIC, the hard band lightcurve appears as the most variable, but conclusions are unclear for the time bins: MOS data show more variability in the smallest time bins, while pn data favor variability in the longest ones. For RGS, the soft band appears as the most variable.

\subsection{Rev. 0903}
A large flare occurred during the last third of the RGS observations. While the EPIC data (esp. pn) are much improved when fitted by an increasing trend, the RGS data are more compatible with a decreasing trend. Both types of instruments agree, however, that the longest time bins generally yield the most variable lightcurves.

\subsection{Rev. 0980}
A large flare occurred near the end of the observation. Data are compatible with a constant, without any trend detected or any obvious oscillation. For EPIC, the largest variability (though not significant) is seen for the total band data while the medium data appear as the most variable in RGS.

\subsection{Rev. 1071}
Only pn data are available for EPIC. There is no localized flare for this exposure but the background smoothly increases in the second half of the observation. An increasing trend is detected for the pn, in the total and soft bands, it is also detected in RGS (esp. RGS1) in the total band, but with a lower significance level. For RGS, the medium band lightcurve appears as the most variable for the longest time bins; for EPIC, the largest variations are detected for shortest time bins in the hard band and the longest time bins in the total band.

\subsection{Rev. 1096}
A few narrow background flares are scattered over the exposure, and a larger one occurs at the end of the observation. Data are compatible with a constant rate, without any trend detected or any obvious oscillation. For RGS, there is no time bin nor energy band favoring variability; for EPIC, the most variable data are found for the longest time bin in the hard band.

\subsection{Rev. 1164}
A few narrow background flares are spread over the exposure, and a larger one occurs at the beginning of the observation. Only MOS data are available for EPIC. A quadratic fit strongly improves the $\chi^2$ in the total and medium bands of the MOS (which are not compatible with a constant count rate, at the 1--10\% level), and in the total band of the RGS. For both types of instruments, the longest time bins provide the most variable lightcurves, but the most variable energy bands are the total and soft bands for MOS and RGS, respectively.

\subsection{Rev. 1343}
A large flare occurred during the first third of the observations. Data in the medium, total and Bergh\"ofer's EPIC bands are incompatible with a constant and significantly better fitted by a linear decrease. The non-constancy in the Bergh\"ofer's band is also detected in the RGS data, though with a lower significance level. For EPIC, the medium band data and lightcurves calculated using 5ks bins appear as the most variable; for RGS, Bergh\"ofer's band appears the most variable, especially at small time bins.

\subsection{Rev. 1620}
A large flare occurred near the end of the observation. Data in all EPIC bands but the hard one are incompatible with a constant and significantly better fitted by a linear increase. The trend in the total band is also detected in the RGS data. For EPIC, the total and medium band data appear as the most variable, especially for the longest time bins; for RGS, the longest time bins yield the most variable lightcurves for the total and Bergh\"ofer's bands.

\subsection{Rev. 1814}
A small flare occurred at the beginning of the RGS observation. Data in all EPIC bands but the hard one are incompatible with a constant and are significantly better fitted by a large linear increase or, even better, a quadratic increase. The non-constancy and the improvement by an increasing trend are also detected in the RGS data for the total and medium bands. For EPIC, the hard data appear as the least variable, while the lightcurves calculated with the longest time bins are the most variable. For RGS, the medium band data are the most variable.

\subsection{Rev. 1983}
A large soft proton flare occurred near the end of the RGS observation. Data in the total and medium EPIC bands are significantly better fitted by a linear, or even quadratic, increase, while the hard and Bergh\"ofer's bands are only better fitted by a quadratic increase. The increasing trend is also detected in the RGS data for the total band. An oscillation is visible in Bergh\"ofer's band on both RGS and EPIC, with a recurrence time of about 50ks. In general, the longest time bins yield the most variable lightcurves.

\end{appendix}

\end{document}